\documentclass[12pt]{emulateapj}
\usepackage{color}
\usepackage{natbib}
\usepackage{amsmath}
\usepackage{times}
\usepackage{graphicx}
\usepackage{epstopdf}
\usepackage{hyperref}

\def\be{\begin{eqnarray}}   \def\ee{\end{eqnarray}}
\def\ben{\begin{equation}\begin{aligned}} \def\een{\end{aligned}\end{equation}}

\def\sec#1{Section~\ref{sec:#1}}
\def\fig#1{Figure~\ref{fig:#1}}
\def\tab#1{Table~\ref{tab:#1}}
\def\equ#1{Equation~\ref{equ:#1}}

\newcommand{\degree}{\ensuremath{^\circ}}
\definecolor{grey}{rgb}{0.35,0.35,0.35}
\begin{document}
\title {Stellar population synthesis based
modelling of the Milky Way using asteroseismology of 13000
{\it Kepler} red giants.}
\author{Sanjib Sharma\altaffilmark{1}, Dennis Stello\altaffilmark{1,2}, Joss
Bland-Hawthorn\altaffilmark{1}, Daniel Huber \altaffilmark{1,2,3}, Timothy R.  Bedding\altaffilmark{1,2}}
\altaffiltext{1}{Sydney Institute for Astronomy, School of Physics,
University of Sydney, NSW 2006, Australia}
\altaffiltext{2}{Stellar Astrophysics Centre, Department of
Physics and Astronomy, Aarhus University, DK-8000 Aarhus
C, Denmark}
\altaffiltext{3}{SETI Institute, 189 Bernardo Avenue, Mountain View, CA 94043, USA}

\begin{abstract}
With current space-based missions it is now possible to
obtain age-sensitive asteroseismic information for
tens of thousands of red giants. This
provides a promising opportunity to study the Galactic
structure and evolution.
We use asteroseismic
data of red giants, observed by {\it Kepler}, to test
the current theoretical framework of modelling the Galaxy
based on population synthesis modeling and the
use of asteroseismic scaling
relations for giants.
We use the open source code {\sl Galaxia} to model
the Milky Way and find the distribution of the masses
predicted by {\sl Galaxia} to be
systematically offset with respect to the seismically-inferred observed masses. The
Galactic model overestimates the
number of low mass stars, and these stars are predominantly
old and of low metallicity.
Using corrections to the $\Delta \nu$ scaling relation
suggested by stellar models significantly
reduces the disagreement
between predicted and observed masses.
For a few cases where non-seismic mass estimates are
available,  the corrections to $\Delta \nu$ also improve the
agreement between seismic and non-seismic mass
estimates.
The disagreement between predictions of the Galactic model
and the observations is most pronounced for stars with
${\rm  [Fe/H]}<-0.5$ and ${\rm  [Fe/H]}>0$ or for
$T_{\rm eff}>4700$ K.
Altering the star formation rate in order to
suppress stars older than 10 Gyr improves the agreement for
mass but leads to inconsistent  color distributions.
We also tested the predictions of the TRILEGAL Galactic model.
However, unlike {\sl Galaxia}, it had difficulties in reproducing the
photometric properties of the Kepler Input Catalog because it
overestimates the number of blue stars.
We conclude that either the
scaling relations and/or the Galactic models need to be
revised  to reconcile predictions of theory with asteroseismic observations.
\end{abstract}

\section{Introduction}
Our understanding of the formation of the Milky Way
is seriously hampered by our inability to reliably
measure ages of stars. In recent years,
asteroseismology has been been demonstrated as
a promising method to estimate fundamental stellar
properties, including ages.
From the seismology one can infer stellar radius and mass,
and for red giants, mass is a good age indicator when the
metallicity is known \citep{2013ARA&A..51..353C}.
The fact that we can access asteroseismic information for a large
number of stars, due to missions like {\it
Kepler} \citep{2011MNRAS.414.2594H,2013ApJ...765L..41S}, {\it CoRoT}
\citep{2010A&A...517A..22M,2013MNRAS.429..423M}, and now also K2
\citep{2015arXiv150608931S}, means that we can start using
this information to
unravel the formation history of the Milky Way.

We can use asteroseismic information to model the Milky
Way within the framework of stellar population
synthesis \citep{1986A&A...157...71R}.
Such models generate a synthetic catalog of
stars for a given prescription of galactic structure
and evolution. The predictions can then be directly compared
to observations. A few such models are publicly available
and have been developed and fine tuned to satisfy
observational constraints from various photometric,
spectroscopic, and astrometric surveys. These include
{\sl Besan\c{c}on} \citep{2003A&A...409..523R}, TRILEGAL
\citep{2005A&A...436..895G} and {\sl Galaxia}
\citep{2011ApJ...730....3S}.

The first study to make use of asteroseismic information to test
a stellar population synthesis-based model was performed by
\citet{2009A&A...503L..21M}.
A more comprehensive
study was carried out by \citet{2011Sci...332..213C}, who
compared the mass and radius
distributions of about 400 main-sequence and subgiant stars
with the {\rm TRILEGAL}
stellar population synthesis model
\citep{2005A&A...436..895G}.
They found that the radius distribution
matched with the model predictions but the mass distribution
showed significant differences.
Specifically, the model was
found to under predict the number of low mass stars ($M<1.15{\rm M}_{\odot}$).
However,
the {\it Kepler} main-sequence/subgiant sample is small in
size,  is local to the Sun and is dominated by young stars, which
limits it applicability  for detailed model comparison.

In comparison to main-sequence and subgiant stars, red giants
offer a number of advantages. For red giants the age is
almost independent of luminosity. Hence,
for any given apparent-magnitude-limited sample, one can
obtain red giants spanning a wide range in age.
Being luminous, they probe the
Galaxy further for any given apparent magnitude and allow
us to sample different regions of the Galaxy.
The ability to sample different regions of the Galaxy is essential for
Galactic archaeology because
different regions of the Galaxy are dominated by
different components. For example,
the fraction of old stars increases with height above the
Galactic mid plane.

\citet{2013MNRAS.429..423M} carried out a differential study
of the Milky Way using
red giants with asteroseismic information from {\it  CoRoT}.
Using about
2000 red giants in two different regions of the Galaxy, they
showed that the mass distributions differ,
in agreement with theoretical models. However, they did not make a direct
comparison with Galactic models.
With {\it Kepler}, we now have a sample of oscillating
red giants  that is
more than an order of magnitude larger than any
other asteroseismic sample that has been
tested against predictions of stellar population synthesis
models \citep[12964 red giants by][]{2013ApJ...765L..41S}.
The aim of this paper is to use this large
sample of giants in the context of modeling the
Galaxy.

However, a significant outstanding problem of using red
giants is that modeling their individual frequencies is too
time consuming for the analysis of tens of thousands of
stars. We therefore rely on using asteroseismic
scaling relations,
$\nu_{\rm max} \propto g T_{\rm eff}^{-1/2}$ and
$\Delta \nu \propto \rho $
\citep{1991ApJ...368..599B,1995A&A...293...87K}, to estimate
their radius and mass (and hence age). Here,
$\nu_{\rm max}$ is the frequency of maximum amplitude and
$\Delta \nu$ the average large frequency separation.
These relations
assume that the structure of a red giant star is homologous with respect to
the Sun. In reality this assumption is not strictly correct
and verification of the relations is ongoing.
However,  independent high-precision
estimates of mass and radius required for this
verification are difficult to obtain.
For subgiants and dwarfs, the $\nu_{\rm max}$ scaling
relation has been shown to work well
\citep{2014aste.book...60B} and recently,
\citet{2015MNRAS.451.3011C} found the proportionality
$\nu_{\rm max} \propto g T_{\rm eff}^{-1/2}$ to be
accurate to within 1.5\%.
Using Hipparcos parallaxes and/or
interferometry, the asteroseismic radii calculated from
scaling relations have been found to be accurate to within 5\%
\citep{2010MNRAS.405.1907B,2012ApJ...757...99S,2012ApJ...760...32H}.
For giants we generally do not have accurate parallaxes, so such studies are
awaiting results from Gaia \citep{perryman02}.
Open clusters have been used to test the
scaling relations for giants \citep{2012A&A...543A.106B,2013ApJ...762...58S,2012MNRAS.419.2077M}. \cite{2012MNRAS.419.2077M} found
agreement to within 5\% for scaling relation-based radii.
Testing of masses is more challenging.
For a few cases where such verification have been performed, the
scaling relation-based masses seem to be overestimated for giants
\citep{2012MNRAS.419.2077M,2013A&A...556A.138F,2014ApJ...785L..28E}.
For two lower red giant branch stars
\citep{2014ApJ...785L..28E} find evidence that
the mass estimated by using only $\Delta \nu$ (but with
additional $\nu_{\rm max}$ independent quantities) is
lower compared to using both $\Delta \nu$ and
$\nu_{\rm max}$. Based on this they suggested
that a modification to the $\nu_{\rm max}$ scaling
relation might be required.
Theoretical modelling has suggested corrections to the
$\Delta \nu$ scaling relation \citep{2009MNRAS.400L..80S,2011ApJ...743..161W,2013EPJWC..4303004M}, but there has been no
comprehensive study to verify the corrections.
In relation to $\nu_{\rm max}$,
\citet{1999A&A...351..582H} and \citet{2008A&A...485..813C} had
suggested theoretically
that $\nu_{\rm max}$ coincides with
the plateau of the damping rate with frequency.
\citet{2011A&A...530A.142B} confirmed this for the
Sun using SoHO GOLF observations.
\citet{1992MNRAS.255..603B} suggested
that  this is caused by a resonance
between the thermal adjustment time of the superadiabatic
boundary layer and the mode frequency, which was also
confirmed by the theoretical study of \citet{2011A&A...530A.142B}
\citep[see also][]{2012sf2a.conf..173B,2013ASPC..479...61B}.
However, there is currently no way to accurately predict
$\nu_{\rm max}$ from theory.

Additionally, there are a few factors related to the
{\it Kepler} red giant sample that make
it difficult to use them for population synthesis-based
modeling. Firstly, knowing the
selection function of the stellar sample is an essential requirement for
such an analysis. However, this function has currently not been
quantified. Secondly, the $g,r,i$ and $z$ band
photometry in the {\it Kepler} Input Catalog
\citep[KIC,][]{2011AJ....142..112B} differs
slightly from the corresponding Sloan magnitudes
\citep{2012ApJS..199...30P}. This means the
synthetic photometry that comes with isochrones, which is calibrated to
Sloan magnitudes, needs to be
re-calibrated, and this has not been done.
We specifically address these issues here.

The paper is organized as follows. Section 2, describes
our methods. We first describe the Galactic models that are
compared against the seismic observations.
Next, we discuss how to map the synthetic stars
of Galactic models into the observational space.
Following this, we discuss how to convert mass and radius to
asteroseismic observables and vice verse.
In Section 3, we analyze the {\it Kepler} giant
samples.
We investigate corrections to
the scaling relations in Section 4, and corrections
to the Galactic model in Section 5.
Finally, in Section 6, we discuss implications
of our findings.

\section{Methods}
\subsection{Stellar-population-synthesis-based modeling of the Milky Way}
The main Galactic stellar-population-synthesis model
used in this paper is from the {\sl Galaxia} code \footnote{\url{http://www.galaxia.sourceforge.net}}
\citep{2011ApJ...730....3S}. It uses a Galactic model
based on the {\sl Besan\c{c}on} model
by \citet{2003A&A...409..523R} but with some modifications.
{\sl Galaxia} uses its own 3D extinction scheme to specify the dust
distribution. We also apply a low latitude correction to the
dust maps as in \citet{2014ApJ...793...51S}.
The isochrones to predict the stellar
properties are from the Padova database
\citep{2008A&A...482..883M,1994AAS..106..275B}. The unique feature of {\sl
Galaxia} is its novel star-spawning scheme which,
unlike previous codes, does not discretize the spatial
dimensions into multiple lines of sight. Instead,  it
generates a continuous three-dimensional distribution of
stars.

Full details of the Galactic model are
available in \citet{2003A&A...409..523R} and
\citet{2011ApJ...730....3S}, but here we summarize the main features.
The Milky Way in {\sl Galaxia}/{\sl Besan\c{c}on} consists
of four major components: the thin
disc, the thick disc, the bar shaped bulge, and the
halo. Each component has its
own initial mass function (IMF) and an analytic formula for the spatial
distribution of stars.
The thin disc is built up using a star-formation history,
while other components are assumed to be populations of
fixed age.
In the thin disc, the metallicity of stars
is governed by an age metallicity relation and the spatial
distribution of stars is governed by an age scale-height
relation. The full {\sl Besan\c{c}on} Galactic model is tuned to
satisfy constraints from the
Hipparcos mission \citep{1997ESASP1200.....E} and star counts from surveys in optical
and near infrared bands \citep{2003A&A...409..523R}. The
Galactic potential is computed in a self-consistent manner
taking into account the results from the Hipparcos mission.

In addition to {\sl Galaxia}, we used the TRILEGAL\footnote{\url{http://stev.oapd.inaf.it/cgi-bin/trilegal}}
Galactic stellar-population-synthesis model
\citep{2005A&A...436..895G}. Unlike {\sl Galaxia}, TRILEGAL
cannot generate stars over a wide angular area. Therefore, we
generated stars along 21 lines of sight pointing towards the
centers of the 21 {\it Kepler} CCD-module foot prints on sky.
We used TRILEGAL with the default settings but with
binary stars turned off.
For extinction, we used
the extinction model of {\sl Galaxia}.
One noticeable difference between {\sl Galaxia} and
TRILEGAL is that, {\sl Galaxia} uses a constant star formation rate,
while the default setting in TRILEGAL
uses a two-step star-formation rate,  in which the rate
between 1-4 Gyr is
twice that at any other time.

\begin{figure}
\centering \includegraphics[width=0.5\textwidth]{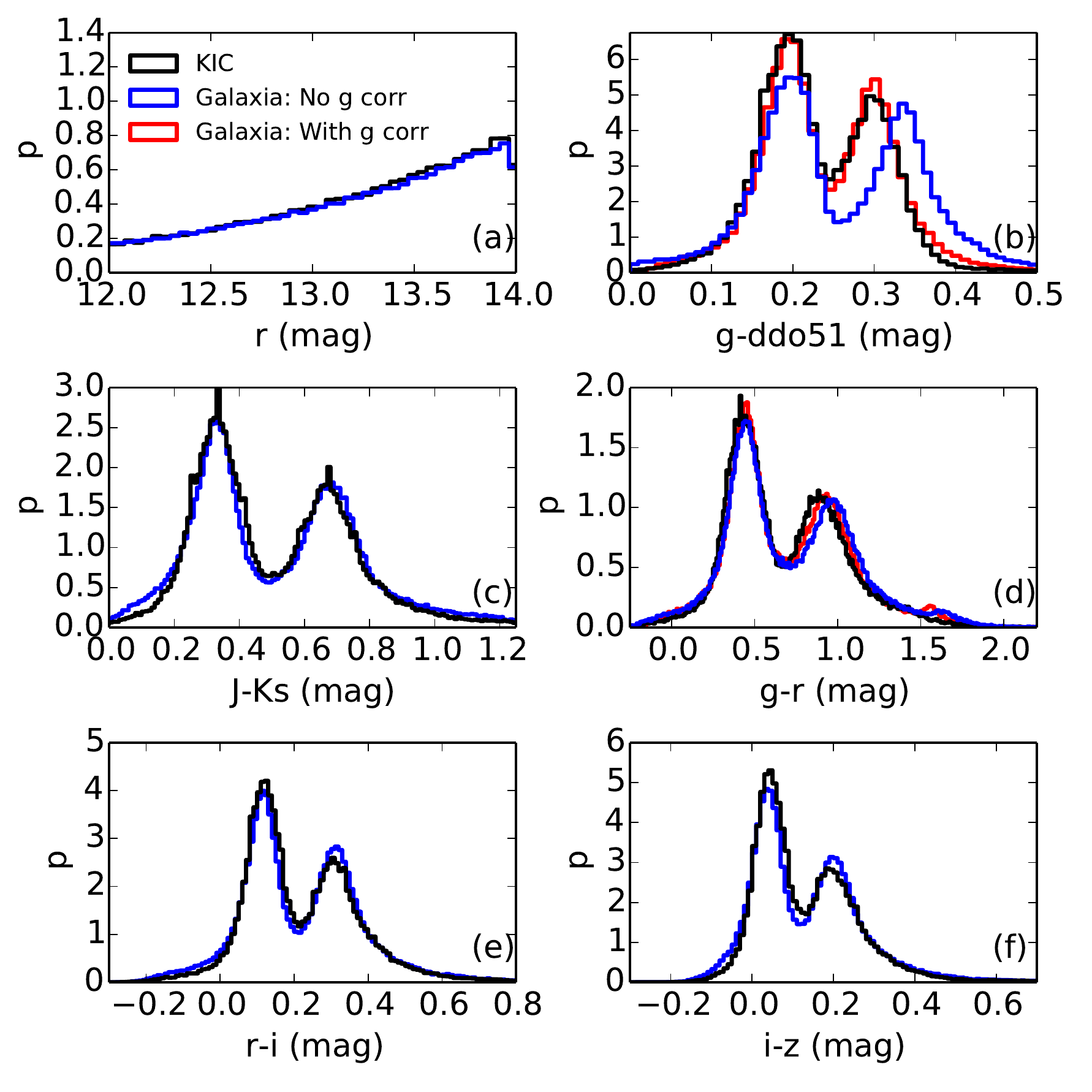}\caption{ Comparison of photometry  between stars in the KIC
and   the synthetic catalog generated by the code {\sl
Galaxia}.
A correction term was applied to the  $g$ band
photometry of the synthetic stars to match the
KIC distribution of the $g-{\rm ddo51}$ color (panel (b)). The blue line is for the
case with no correction. The correction also improves the
fit to the $g-r$ color distribution, in regions near the
red peak (panel (d)). $p$ stands for probability density whose
integral over the abscissa is unity.
\label{fig:kic_photo1}}
\end{figure}

\subsubsection{Comparing the synthetic catalog photometry to the {\it
Kepler} input catalog}  \label{sec:comp_photo}
Before we can start comparing the asteroseismic results with predictions
from our Galactic models, we have to make
sure that the photometry of stars in the KIC and
the Galactic model predictions agree with each other.
This is because the
asteroseismic targets observed by {\it Kepler} were
selected based on KIC parameters, which ultimately relied
in its photometric entries.
If the photometry of the Galactic model does not match with the
KIC,  the model will be fundamentally
inconsistent with observations, preventing meaningful model-based
inferences to be made.
To proceed, we first
compared the photometry of the synthetic stars with that of
stars in the KIC and identified any differences.
Next, we investigated the cause of any mismatch
and derive transformation formulae to rectify it.

\begin{figure}
\centering \includegraphics[width=0.5\textwidth]{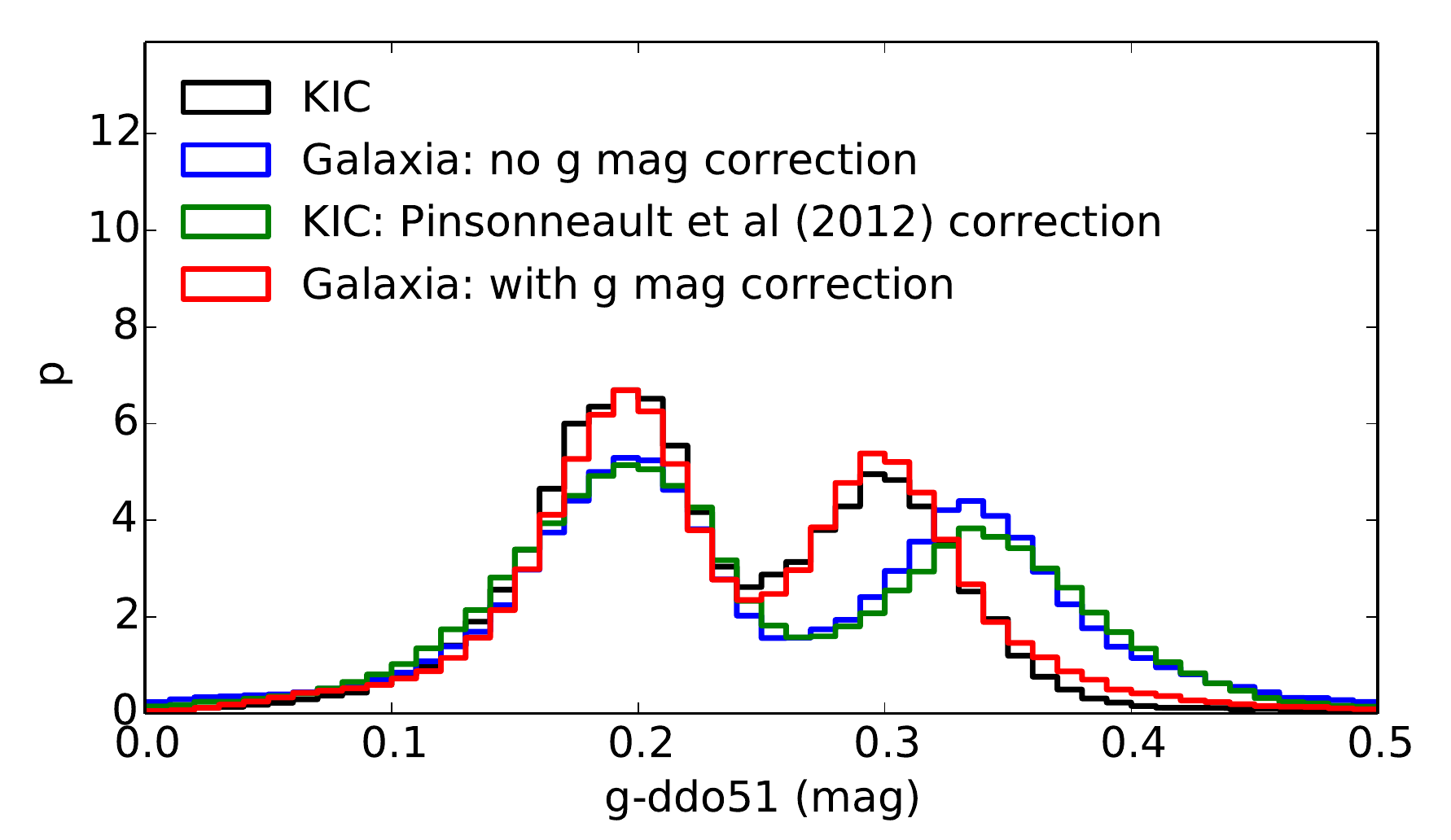}\caption{ Comparison of the $g-{\rm ddo51}$ distribution of stars in the
KIC (black) with
the synthetic catalog generated by the code {\sl
Galaxia} (blue). The green and red distributions show
the effect of different
transformation formulae to convert between KIC and SDSS
$g$ band magnitudes, with results of the
\citet{2012ApJS..199...30P} correction applied
(Equation-\ref{equ:pinson_tr}) to KIC
stars (from black to green), and results of our correction
(Equation-\ref{equ:gkic}) applied to stars in the synthetic
catalog (from blue to red).
\label{fig:kic_photo2}}
\end{figure}

We generated a synthetic catalog of
the Milky Way using the code {\sl Galaxia}. Stars,
both from the synthetic catalog and from the KIC,
that lie within 8 degrees from the
center of the {\it Kepler} field and with magnitude $r<14$ were
selected for comparison.
Note the KIC is expected to be complete to magnitudes even
fainter than 14 magnitude in $r$.
We then added photometric
errors to the stars.
For this we derived the following formulas to
approximate the actual photometric errors in 2MASS
and the KIC \citep[Figure 4 in][]{2011AJ....142..112B}:
\be
\sigma_{\rm 2MASS} & = & 0.0125+0.01\exp(K-12.6)\\
\sigma_{\rm SDSS} & = & 0.015+0.01\exp(r-16.0),
\ee
In \fig{kic_photo1}, we show the distribution of $r$ band
magnitude and various colors for both the KIC (black) and the
synthetic catalog (blue) \footnote{In this paper we plot
normalized distributions
such that the integral over the abscissa is
unity. This corresponds to probability density and is
labelled as $p$.}.
The $r$ magnitude and the color distributions match
well, which is encouraging given that we do not enforce
any fine tuning of the model.
However, the distributions
of $g-{\rm ddo51}$ and $g-r$ (\fig{kic_photo1}b,d)
for the synthetic catalog are slightly redder for the giants
(right most peaks).
The discrepancy for $g-{\rm ddo51}$ is especially large.
Either the $g$ band or  the ${\rm ddo51}$ band photometry,
or both could be discrepant.
A clue to the cause of the discrepancy is provided by
results of \citet{2012ApJS..199...30P}, who found
systematic differences in the KIC $(u,g,r,i,z)$ photometry
with respect to SDSS $(u,g,r,i,z)$ photometry, for a small
part of the {\it Kepler} field which overlaps with the SDSS.
Because the isochrones that we use are
calibrated to the SDSS bands, this could partly
explain the mismatch that we see with the KIC.
To investigate this, we show in \fig{kic_photo2}, a copy of
\fig{kic_photo1}b, but we add (in green) the $g-{\rm ddo51}$
color  after applying  the \citet{2012ApJS..199...30P}
transformations to the KIC stars.
We see that this corrected KIC distribution
matches the predictions of {\sl Galaxia} (blue).
Hence, a correction in $g$ band is enough to explain the
mismatch seen in the distribution of  $g-{\rm ddo51}$
color.
The transformations proposed by \citet{2012ApJS..199...30P}
were:
{\small
\begin{equation}
\label{equ:pinson_tr}
\begin{aligned}
& g_{\rm SDSS}  =  g_{\rm KIC}+0.0921(g_{\rm KIC}-r_{\rm
KIC})-0.0985  +[0.055]  \\
& r_{\rm SDSS}  =  r_{\rm KIC}+0.0548(r_{\rm KIC}-i_{\rm
KIC})-0.0383  +[0.0]    \\
& i_{\rm SDSS}  =  i_{\rm KIC}+0.0696(r_{\rm KIC}-i_{\rm
KIC})-0.0583  +[0.02]   \\
& z_{\rm SDSS}  =  z_{\rm KIC}+0.1587(i_{\rm KIC}-z_{\rm
KIC})-0.0597  +[0.02].
\end{aligned}
\end{equation}
}
A slight
adjustment of zero-points (see square brackets) from the
original \citet{2012ApJS..199...30P}
transformations was needed to make the green curve match
with the {\sl Galaxia} prediction. Note slight
adjustments of zero points are quite common when
comparing photometry obtained from different sources.

The transformations of \citet{2012ApJS..199...30P}
convert KIC photometry to SDSS.
However, for our analysis we require the
inverse transformation, that is from SDSS to KIC colors.
This is because the stars observed by {\it Kepler}
were selected from the KIC photometry and not a re-calibrated version of
it. Equations \ref{equ:pinson_tr} can be written in a matrix
form and then inverted to get the
transformation formula for $g_{\rm KIC}$, but this would require
knowing the magnitudes in three bands $g_{\rm SDSS},r_{\rm SDSS}$ and
$i_{\rm SDSS}$.
Instead, for our use, we derived a simpler formula
\be
g_{\rm KIC}=g_{\rm SDSS}-0.25(g_{\rm SDSS}-{\rm
ddo51})+0.048
\label{equ:gkic}
\ee
that makes use of only two bands, $g_{\rm SDSS}$ and ${\rm
ddo51}$. The formula is easily invertible to
$g_{\rm    SDSS}=g_{\rm KIC}+(g_{\rm KIC}-{\rm
ddo51})/3-0.064$. Moreover, because  $g_{\rm SDSS}$ and
${\rm  ddo51}$ are on different photometric systems,
their errors are less likely to be correlated, which
means the formula given by \equ{gkic} is more robust.
The formula was derived  by matching
the  $g_{\rm KIC}-{\rm ddo51}$
distribution of synthetic stars
to those in the KIC.
The red curve in \fig{kic_photo2} shows the results after applying
\equ{gkic} to the $g$ band photometry of the synthetic catalog, which makes
the $g-{\rm ddo51}$ color  distribution of synthetic stars
agree nicely with that of the KIC (black). This result is
copied onto \fig{kic_photo1}b. The formula was also found to
improve the agreement for the $g-r$
color distribution, which is shown in \fig{kic_photo1}d.
Thus, we can now successfully reproduce the photometry
of the KIC stars using {\sl Galaxia}.
For the rest of the paper we use
the above formula to transform the $g$ band photometry
of synthetic stars to compare the
results of {\sl Galaxia} with observations.
In general, the formula should be valid for the version
of Padova isochrones adopted in {\sl Galaxia}.
We note that \citet{2013MNRAS.433.1133F} has also
suggested transformations from SDSS to KIC photometry,
but those transformations (like ours) are specific to the
adopted stellar
models and to the adopted bolometric corrections.
We found their transformations to be inadequate to explain the
mismatch of the $g-{\rm  ddo51}$ color that we encounter.

\subsection{Estimating KIC stellar parameters of stars in
the synthetic catalog}
\label{sec:scp}
As mentioned in the previous section,
stellar targets of the {\it Kepler} mission were selected
based on stellar parameters in the KIC, such as
$T_{\rm eff}, \log g, \log Z, M$, and $R$.
These parameters were derived from photometry
as part of the stellar classification project (hereafter
SCP). As part of the SCP, \citet{2011AJ....142..112B} developed a software
code\footnote{https://www.cfa.harvard.edu/kepler/kic/kicindex.html}
to generate stellar parameters
for the stars in the KIC based on  a Bayesian posterior
maximization scheme. In order to reproduce the
selection function of the observed targets,
we used this SCP software code
to generate KIC-equivalent stellar parameters for stars in
the synthetic catalog.

To run the SCP code, one has to specify the
photometric bands to be used and the photometric
uncertainty in each band.
While the KIC contains photometry
in 10 bands, we chose to use only the
following eight $(g, r, i, z, J, H, K_s, {\rm
D51})$ that are available for the majority of stars.
We set the photometric uncertainty to 0.005 which was found
to best reproduce the KIC stellar parameters when applied
to KIC stars.  The adopted photometric uncertainty
is close to but smaller than the quoted typical uncertainty in
the KIC of $\sigma=0.0175$ mag for $r<14$
\citep{2011AJ....142..112B}.

\begin{figure}
\centering \includegraphics[width=0.5\textwidth]{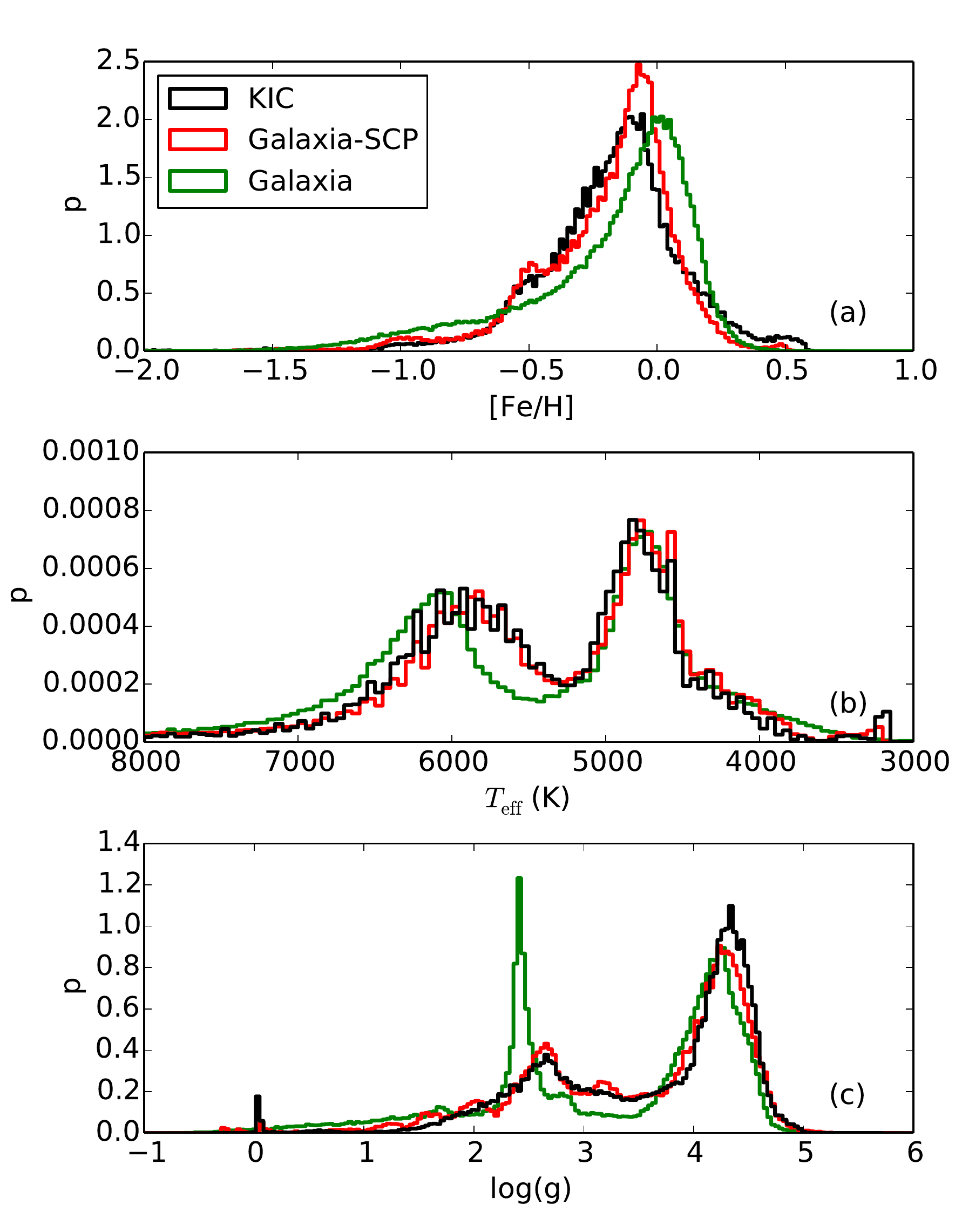}\caption{Comparison between observed (black) and predicted (using {\sl Galaxia})
distributions of stellar parameters for stars in the KIC.
The red line shows the stellar parameters of synthetic stars
estimated with the SCP code, which agrees well with the black
line of KIC stars. The green line shows the true
stellar parameters of the synthetic stars. Comparing the
red and green lines reveals that
the SCP underestimates the metallicity
(panel (a)) and
the temperature  of hot stars (panel (b)).
Panel (c) shows that the
sharp peak in gravity of red clump stars is not well reproduced by the
SCP code.
\label{fig:kic_spc2}}
\end{figure}

Finally, we checked if our population synthesis model
can match the distribution of stellar parameters in the
KIC. The comparison is shown in \fig{kic_spc2}.
We see that the SCP-generated stellar parameters
in the synthetic catalog (red line) match well with that
of the KIC (black line), which is a prerequisite
to reproduce the selection function
of the {\it Kepler} mission.
As an aside, we compared the true
stellar parameters (green line) of the synthetic catalog with that of
their SCP estimates (red line). We see slight differences.
The SCP code underestimates
${\rm [Fe/H]}$ by about 0.1 dex.
The gravity of giants ($\log g < 3.5$) is overestimated and
also has significant uncertainty
(red peak at $\sim 2.6$ is broadened and shifted).
Finally, for stars hotter than 5200 K
SCP underestimates $T_{\rm  eff}$ by about 300 K.
This means that the KIC underestimates the
radii of the main sequence stars.
A similar effect was also reported by
\cite{2011ApJ...738L..28V} while
comparing the asteroseismic
radii with the KIC radii for a sample of
subgiants and dwarfs.
The underestimation of radius by KIC, has important
implcations for detecting planets in habitable zones
around these stars.
Specially, this would lead to underestimation of planet
radii when estimated using the transit method.
However, this has no implication for the analysis of the
red giant stars presented in this paper.

\begin{figure}
\centering \includegraphics[width=0.5\textwidth]{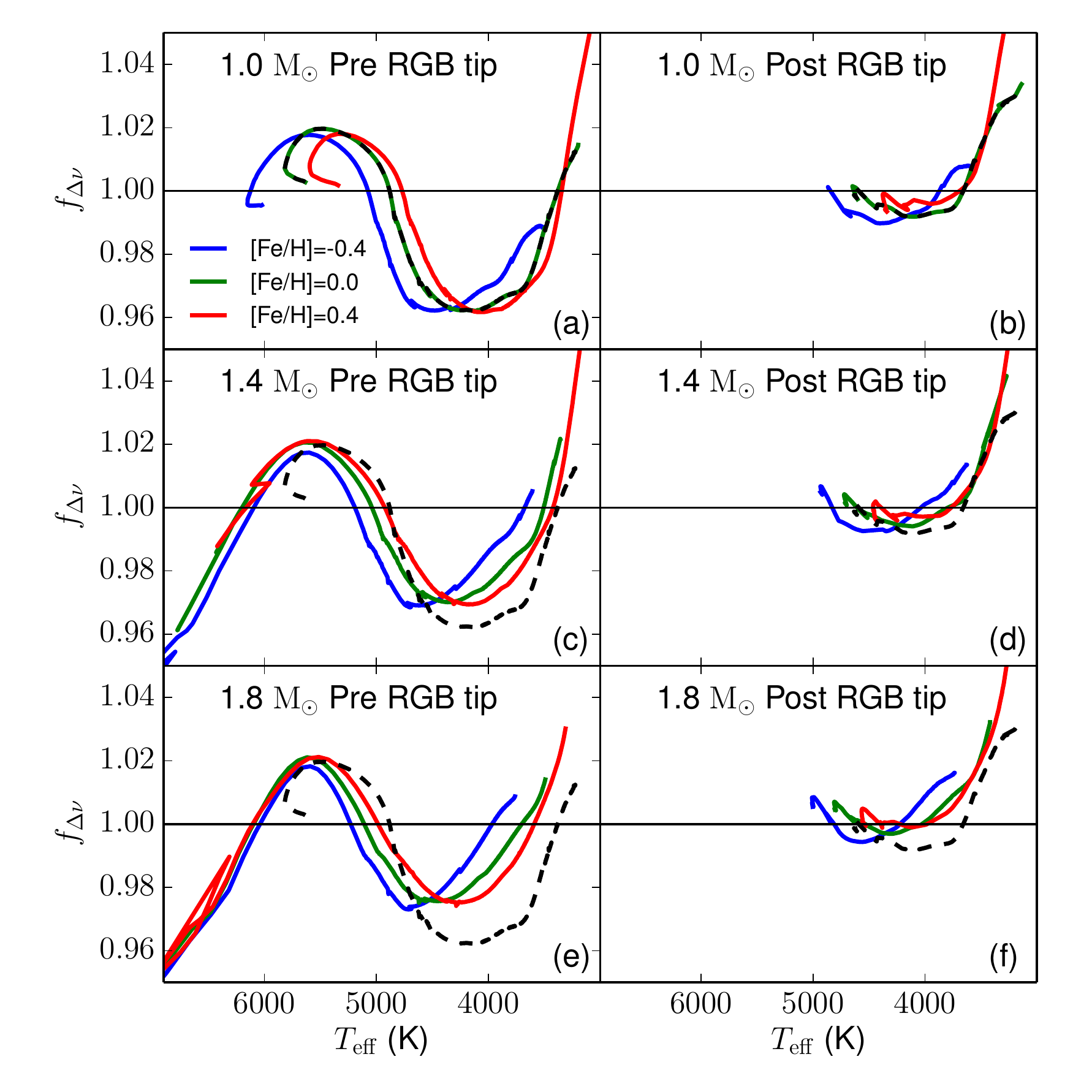}\caption{The correction factor $f_{\Delta \nu}$  predicted
by stellar models as a
function of $T_{\rm eff}$ for  different  metallicity,
mass, and evolutionary state.
Each row corresponds to
a different mass labelled on the panels.
Left panels: evolution from the main sequence till the tip of the
red giant branch. Right panels: evolution from
the onset of stable helium-core burning to the start of the
asymptotic giant branch (end of helium-core burning).
The dashed reference line is for the case with
${\rm [Fe/H]}=0.0$ and
$M=1.0 {\rm M}_{\odot}$. We compute ${\rm [Fe/H]}$ assuming $Z_{\odot}=0.019$.
\label{fig:teff_dnu_grid}}
\end{figure}

\subsection{Asteroseismic scaling relations and estimation of mass and radius}
\label{sec:massrad}
Cool stars exhibit convectively driven oscillations whose frequency spectra
show a pattern of peaks similar to that seen for the Sun.
The near
regular pattern is characterized by the so-called large frequency
separation, $\Delta \nu$, between overtone modes. The amplitude of the peaks
is modulated by an envelope, which has a central
frequency of maximum amplitude, $\nu_{\rm max}$.
Theory indicates that $\Delta \nu$ is related to the density
of the star \citep{1986ApJ...306L..37U}, while $\nu_{\rm
max}$ has been conjectured to
depend upon the acoustic cut-off frequency of the atmosphere,
and hence the surface gravity and the effective temperature of the star
\citep{1991ApJ...368..599B,1995A&A...293...87K, 2011A&A...530A.142B}.
This leads to the following approximate scaling relations:
\be
\label{equ:scaling_freq1}
\frac{\Delta
\nu}{\Delta \nu_{\odot}}  & \approx &
f_{\Delta \nu} \left(\frac{\rho}{\rho_{\odot}}\right)^{0.5},
\\
\label{equ:scaling_freq2}
\frac{\nu_{\rm max}}{\nu_{\rm max, \odot}}  &  \approx  &
f_{\nu_{\rm max}}\frac{g}{{\rm g}_{\odot}} \left(\frac{T_{\rm eff}}{T_{\rm eff,
\odot}}\right)^{-0.5}.
\ee
Here, $f_{\Delta \nu}$ and $f_{\nu_{\rm max}}$ are
correction factors, which we have introduced to quantify any
deviation from the scaling relations.
The correction factors
are degenerate with the choice of solar reference values.
We fix this by adopting a consistent definition for
the solar reference values. Throughout the paper
the solar reference values that we use are
, $\Delta \nu_{\odot}=135.1\ \mu
{\rm Hz},\
\nu_{\rm max,\odot}=3090\ \mu {\rm Hz}$, and $T_{\rm
eff,\odot}=5777 {\rm \ K}$ and are adopted from
\citet{2011ApJ...743..143H, 2013ApJ...767..127H}.
These are the values of the Sun obtained using the
\citet{2009CoAst.160...74H} pipeline, which was used by
\citet{2013ApJ...765L..41S}.
It is common practice to use such ``method-specific'' solar values, meaning the
values returned from solar data when using the same method as used for the
rest of the stellar sample.  This practice is based on the assumption that
method-specific systematic differences are the same for all
stars \citep[however see][]{2013A&A...556A..59H}.

By rearranging the above equations, one can estimate mass and
radius of a star from the above seismic observables and
effective temperature as follows:
\be
\label{equ:scaling_m}
\frac{M}{\rm {\rm M}_\odot} &  \approx &  \left(\frac{\nu_{\rm
max}}{f_{\nu_{\rm max}} \nu_{\rm max,
\odot}}\right)^{3}\left(\frac{\Delta \nu}{f_{\Delta \nu}
\Delta \nu_{\odot}}\right)^{-4}\left(\frac{T_{\rm eff}}{T_{\rm
eff, \odot}}\right)^{1.5}   \\
\label{equ:scaling_r}
\frac{R}{\rm R_\odot}  & \approx & \left(\frac{\nu_{\rm
max}}{f_{\nu_{\rm max}} \nu_{\rm max, \odot}}\right)\left(\frac{\Delta
\nu}{f_{\Delta \nu} \Delta \nu_{\odot}}\right)^{-2}\left(\frac{T_{\rm
eff}}{T_{\rm eff, \odot}}\right)^{0.5}.
\ee
The temperatures of most of the stars observed by {\it
Kepler} were
estimated from photometry and have significant
uncertainty and systematics that are not
fully understood.  In the following, we therefore
investigate the particular combinations of $\nu_{\rm max}$
and $\Delta \nu$ that comprise the seismic part of
Equations \ref{equ:scaling_m} and \ref{equ:scaling_r}.
This will allow us to quantify the degree of error
introduced into estimating mass and radius from $\Delta
\nu$ and $\nu_{\rm max}$ alone.
Hence, we define the following two quantities,
\be
\label{equ:scaling_m_proxy}
\kappa_M & = & \left(\frac{\nu_{\rm
max}}{\nu_{\rm max,
\odot}}\right)^{3}\left(\frac{\Delta \nu}{  \Delta
\nu_{\odot}}\right)^{-4} \\
\kappa_R & = & \left(\frac{\nu_{\rm
max}}{\nu_{\rm max, \odot}}\right)\left(\frac{\Delta
\nu}{\Delta \nu_{\odot}}\right)^{-2}
\label{equ:scaling_r_proxy}
\ee

To summarize, our strategy to compare asteroseismic
parameters  of synthetic stars with that of observations is as
follows. We know $M$ and $R$ of the synthetic stars, which we use
to estimate $\Delta \nu$ and $\nu_{\rm max}$ using the
scaling relations (Equations \ref{equ:scaling_freq1} and
\ref{equ:scaling_freq2}).
We then compute $\kappa_M$ and $\kappa_R$
of both the
synthetic and observed stars (Equations
\ref{equ:scaling_m_proxy} and \ref{equ:scaling_r_proxy})
and compare them.

The accuracy of the  scaling relations is still not fully
quantified. While one can
predict $\Delta \nu$
from theoretically derived oscillation frequencies, it is
currently not possible to do so for $\nu_{\rm max}$.
In the next subsection
we explore the corrections to the $\Delta \nu$ scaling
relation.

\subsection{Correction to $\Delta \nu$ scaling relation using stellar models}
\label{sec:fdnu_corr}
Analysis
of the theoretical oscillation frequencies derived from stellar
models predict that $\Delta \nu$ does not perfectly scale as
$\rho^{0.5}$
\citep{2009MNRAS.400L..80S,2011ApJ...743..161W,2013EPJWC..4303004M,2013A&A...550A.126M}.
To quantify this deviation we use the correction factor
\be
f_{ \Delta \nu}=\left(\frac{\Delta \nu}{135.1\mu {\rm Hz}}\right)\left(\frac{\rho}{\rho_{\odot}}\right)^{-0.5}.
\ee
The value $135.1 \mu$Hz corresponds to our adopted
choice for $\Delta \nu_{\odot}$.
The stellar models
show that $f_{ \Delta \nu}$
varies with metallicity, $Z$, mass, $M$, and age,  $\tau$.
To  estimate $f_{\Delta \nu}$ we ran a suite of stellar
models for $-1.28<\log Z<2.12$ (corresponding to $-3
<$ [Fe/H] $< 0.4$) and $0.8<M/{\rm
{\rm M}_\odot}<4.0$.
For this we used the MESA (v6950)
\texttt{1M\_pre\_ms\_to\_wd} test suite case
\citep{2011ApJS..192....3P,2013ApJS..208....4P,2015ApJS..220...15P},
but without rotation or mass loss invoked.
We followed the approach by \citet{2011ApJ...743..161W}
to derive $\Delta \nu$ for each model, designed to mimic the
way $\Delta \nu$ is measured from the data. For this we used
GYRE \citep{2013MNRAS.435.3406T} to calculate the required
radial mode frequencies.
An in-depth description of this model grid and some of its
applications will be presented in future
(Stello et al. 2016 in preparation) \footnote{The stellar models along with a code to
compute the correction factors are available at
{\scriptsize \url{http://www.physics.usyd.edu.au/k2gap/Asfgrid}}.}.
We estimated $f_{\Delta \nu}$ along each stellar track
ranging from the zero-age main sequence till the
end of helium-core burning.
The results were remeshed in a grid
in ${\log Z, M,E_{\rm state}, T_{\rm eff}}$ space
with dimensions of {$13\times19\times2\times200$}.
The $E_{\rm state}$ is the evolutionary state, with 0 for pre
helium-ignition evolution and 1 for post helium-ignition evolution.
The age, ${\tau}$, is replaced by variables $E_{\rm
state}$ and $T_{\rm eff}$, which are observable \citep{2011Natur.471..608B}.
Then, using
interpolation we computed the correction factor
$f_{\Delta \nu}$ for each synthetic star and
used it to correct $\Delta \nu$.
From now on we denote this derived correction factor by
$f^{\rm Grid}_{\Delta \nu}$.
We show $f^{\rm Grid}_{\Delta \nu}$ in \fig{teff_dnu_grid} for
different mass and metallicity.
Red giant branch stars require a larger relative correction
(left panels, $T_{\rm eff}<5000$ K) compared to the red clump stars (right
panels). For the red giant branch stars,
the correction factor has a strong dependence on
both temperature and metallicity
\citep[see also][]{2011ApJ...743..161W}. There is also a dependence
on mass.
This can be seen as a shift of the green curves  in panel (c)
and (e) with respect to the dashed curve, which is the
$M=1.0{\rm\ M_\odot}$ equivalent from panel (a).

\section{Analysis of giants}
In this section we study a sample of 12964 red giants from
\citet{2013ApJ...765L..41S}. The observations were obtained
in {\it Kepler}'s long cadence mode (29.4 min).
We first determine the selection
function of the sample and then use that to select
stars from a synthetic catalog generated
using a stellar population synthesis model of the Milky Way.
Next, we compare the properties of the observed stars
with those in the synthetic catalog.

\begin{figure}
\centering \includegraphics[width=0.5\textwidth]{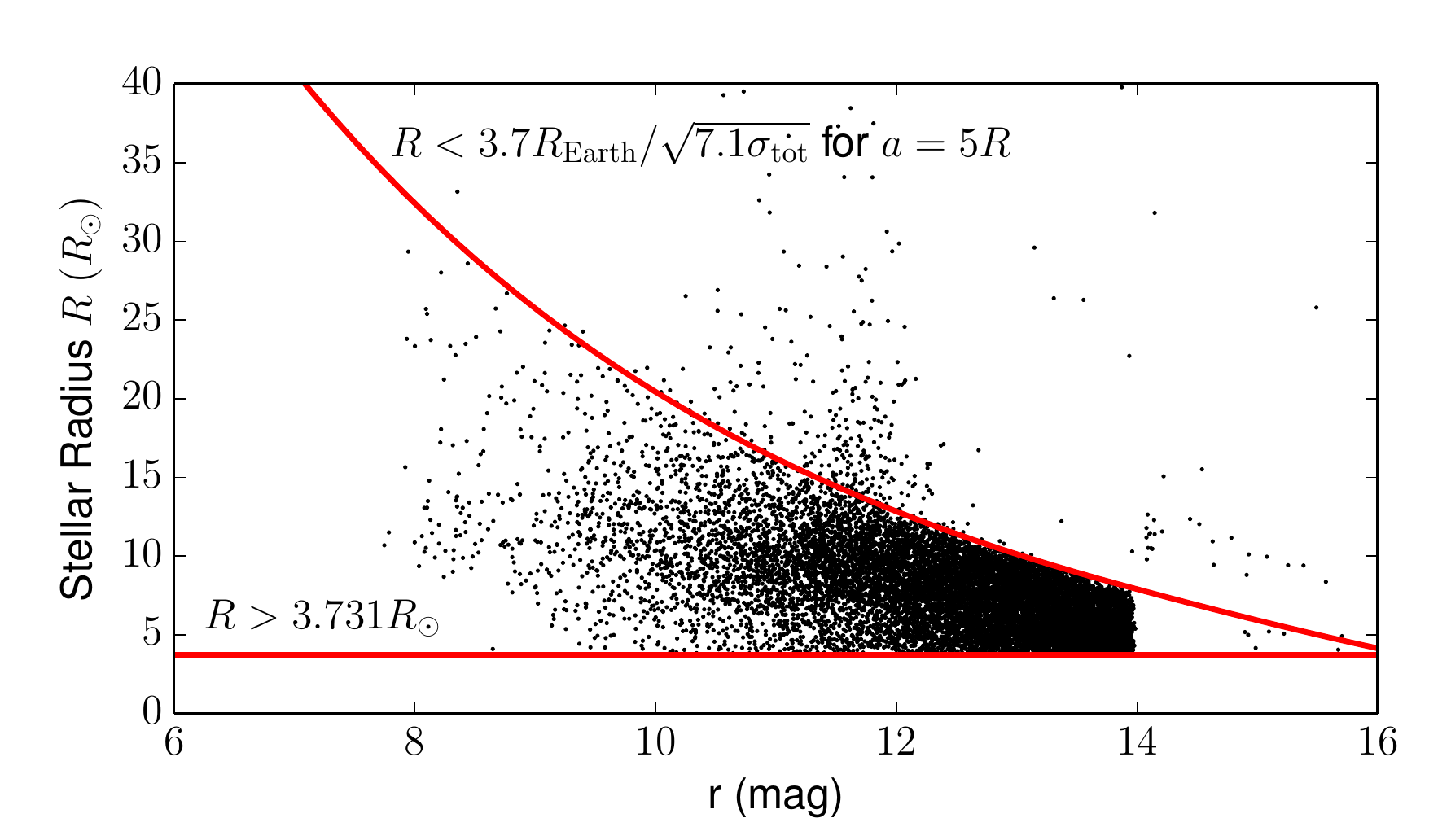}\caption{Selection function of the {\it Kepler} red giant
sample. The plot shows the stellar radius from the KIC as a
function of $r$ band magnitude, which suggests a
selection function of the form $3.731
{\rm\ R}_{\odot}<R<3.7R_{\rm Earth}/\sqrt(7.1\sigma_{\rm
tot})$  (red curves). The upper envelope comes from a limit on the
minimum detectable planet radius assuming a planet orbit
with semi major axis $a=5R$ (see text in \sec{selfunc}). The lower limit is because
the sample was restricted to $\log g_{\rm KIC} < 3.45$.
\label{fig:sel_func_giants}}
\end{figure}

\begin{figure*}
\centering \includegraphics[width=0.98\textwidth]{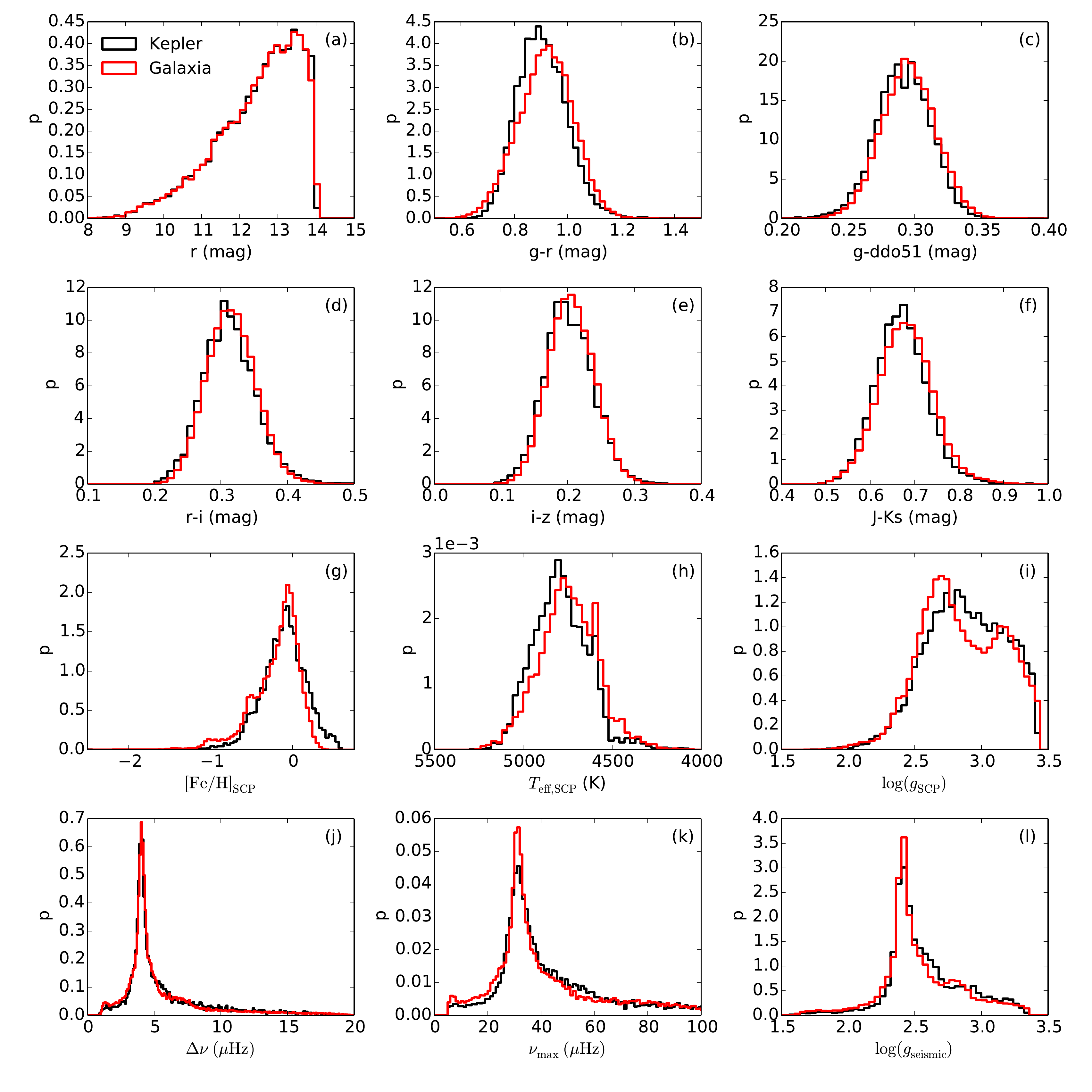}\caption{Comparison of observed and predicted (by {\sl
Galaxia}) distribution of stellar properties for stars in the {\it
Kepler} red giant sample for (a) r band magnitude, (b-f)
various colors, (g-i) stellar
parameters, and (j-l) seismic parameters.
The selection function is given by
$p(r|3.731
{\rm\ R}_{\odot}<R<3.7 R_{\rm Earth}/\sqrt{7.1\sigma_{\rm tot}})$
and was used to sample the stars in the synthetic catalog.
The almost perfect match
in $r$ band is by construction, while a good match in other
distributions, e.g color and stellar parameters, is a
verification of the correctness of the applied selection
function.
\label{fig:giants_summary}}
\end{figure*}

\subsection{The selection function of the {\it Kepler} giants}
\label{sec:selfunc}
In order to compare
observations with Galactic model predictions,
it is crucial that we know the selection
function of the observed targets. However,
so far it has not been possible to reproduce the selection
function that generates the red giant sample observed by
the {\it Kepler} mission. Previous attempts to
characterize it can be found in
\citet{2013MNRAS.433.1133F} and \citet{2014ApJS..215...19P}. The main
source of information is the paper
by \citet{2010ApJ...713L.109B}, which
describes the selection and prioritization of
{\it Kepler} targets in general. The targets were classified into 13
priority groups based on their stellar parameters.
The idea was to prioritize stars for which it was possible
to detect exoplanets, and if possible terrestrial
planets in habitable zones. The compiled list
consisted of  261,636 stars, of which the top 150,000 stars
were initially selected for observations. This included
5282 giants with $\log g_{\rm KIC}<3.5$ and $r<14.0$,
but in the end 17471 stars with these $\log g_{\rm KIC}$ and
$r$ limits were observed. In other
words, more than 50\% of
the total {\it Kepler} red giant sample came from stars that
were added later on by the mission and for which the exact selection
procedure is not documented. Other minor factors that complicate
the selection function are as follows.
Numerous additional targets were observed through the
proposals by the Guest Observer program and the {\it Kepler}
Asteroseismic Science Consortium. Each
of these had their own complex and often undocumented selection schemes.
Moreover, all stars were not observed for the same number of
quarters. The length of observation puts a limit on the
range of $\log g$ over which oscillations can be detected and characterized
\citep[e.g., ][]{2015arXiv150608931S}.

In the following, we deduce an approximate form for the selection function of
the {\it Kepler} giants by comparing the properties of the observed giant sample
against that of all stars in the KIC.
In \fig{sel_func_giants} we plot the KIC radius against the $r$
band magnitude of stars in the KIC.
The majority of stars obey a selection of the
form $(R_{\rm low}<R<R_{\rm up}(r))$, where $R$ is the
stellar radius.
The lower bound $R_{\rm low}=3.731 {\rm R}_{\odot}$ comes from the fact
that the seismic analysis by \citet{2013ApJ...765L..41S} only included
stars with $\log g_{\rm KIC} < 3.45$.
In the KIC,  $\log g_{\rm KIC}$ and $R$
almost follow a one-to-one relation, hence a limit on
$\log g_{\rm KIC}$ implies a limit on $R$.
We found that $R_{\rm up}(r)$ was well described by
(see \fig{sel_func_giants})
\be
R_{\rm up}(r)=\frac{3.7 R_{\rm
Earth}}{\sqrt{7.1\sigma_{\rm LC}/55.37}}
\label{equ:rupper}
\ee
where,  $\sigma_{\rm LC}$ is the long cadence noise given by
\be
\sigma_{\rm LC}=\frac{1}{c}\sqrt{c+7\times10^6\max(1,Kp/14)^4}
\ee
with $c= 3.46\times10^{0.4(12-Kp)+8}$ \citep{2010ApJ...713L.120J}.

The derivation of Equation \ref{equ:rupper}
follows closely the
scheme used by \citet{2010ApJ...713L.109B} to
select and prioritize the planet search targets.
The minimum detectable radius of a planet $R_{\rm planet,min}$ is
given by
\be
R_{\rm planet,min}=R\sqrt{7.1 \sigma_{\rm tot}},
\label{equ:rplanet_min}
\ee
where the noise
\be
\sigma_{\rm tot}=\frac{\sigma_{\rm LC}}{\sqrt{N_{\rm
transit} N_{\rm sample}}}
\ee
depends on the number of transits,  $N_{\rm transit}$,
during the lifetime of the mission, $t_{\rm mission}$,  and
the number of photometric measurements
per transit, $N_{\rm sample}$, with cadence interval, $t_{\rm
cadence}$. $N_{\rm transit}$ and $N_{\rm sample}$
depend on mass, $M$, and radius, $R$, of the host star and on the
semi-major axis, $a$, of the planet orbiting around it, as
follows:
\be
N_{\rm sample}=\frac{4}{\pi}\frac{2R}{t_{\rm cadence}} \sqrt{\frac{a}{GM}} \\
N_{\rm transit}=\frac{t_{\rm mission}}{2 \pi a}\sqrt{\frac{GM}{a}}.
\ee
If the orbit radius is, say, $a=5R$, then for $t_{\rm
mission}=$ 3.5 yrs and $t_{\rm cadence}=$ 30 minutes we have
\be
\sigma_{\rm tot}=\frac{\sigma_{\rm LC}}{55.37}.
\ee
The condition that {\it Kepler} should be able to detect
planets with radius greater than some predefined limit
$R_{\rm planet,lim}$ means,
\be
R\sqrt{7.1 \sigma_{\rm tot}} &< & R_{\rm planet,lim},
\ee
using \equ{rplanet_min}. This puts the following limit on the radius of
the host star
\be
R <\frac{R_{\rm planet,lim}}{\sqrt{7.1\sigma_{\rm tot}(Kp)}}.
\ee
We found that $R_{\rm planet,lim}=3.7 R_{\rm Earth}$ provides
a good match to the upper envelope of the stellar radius
for the red giant sample
(\fig{sel_func_giants}).

\subsection{Comparison of photometric parameters}
Before investigating whether the
mass and radius distributions inferred from asteroseismic
observations match the Galactic model predictions, we first
needed to verify that we have correctly reproduced the
selection function. This was done by comparing the
photometric properties of the two samples.
In \fig{giants_summary} we show the distributions
of $r$ band magnitude, colors, SCP-derived stellar parameters,
the seismic observables $(\Delta \nu$, $\nu_{\rm max})$, and
surface gravity of our giant sample, alongside synthetic
stars from the Galactic model.
The seismic observables for synthetic stars
were calculated using asteroseismic scaling relations
(Equations \ref{equ:scaling_freq1} and
\ref{equ:scaling_freq2} with $f_{\Delta \nu}=f_{\nu_{\rm    max}}=1$).
As before, the perfect
match for the $r$ band distribution (\fig{giants_summary}a)
is by construction, but the good match of the color distributions
(\fig{giants_summary}b-f) and the other parameters gives
confidence that the adopted selection function is correct.
The SCP-derived stellar parameters also match well (\fig{giants_summary}g-i).
The slight difference in the SCP-derived gravity distribution
is due to our inability to exactly reproduce SCP-derived
gravity,  as discussed in \sec{scp}.
Finally, the distribution of the seismic parameters
$(\Delta \nu$, $\nu_{\rm max})$, and the inferred gravity
also agree with the predictions of the
Galactic model (\fig{giants_summary}j-l).

\begin{figure}
\centering \includegraphics[width=0.5\textwidth]{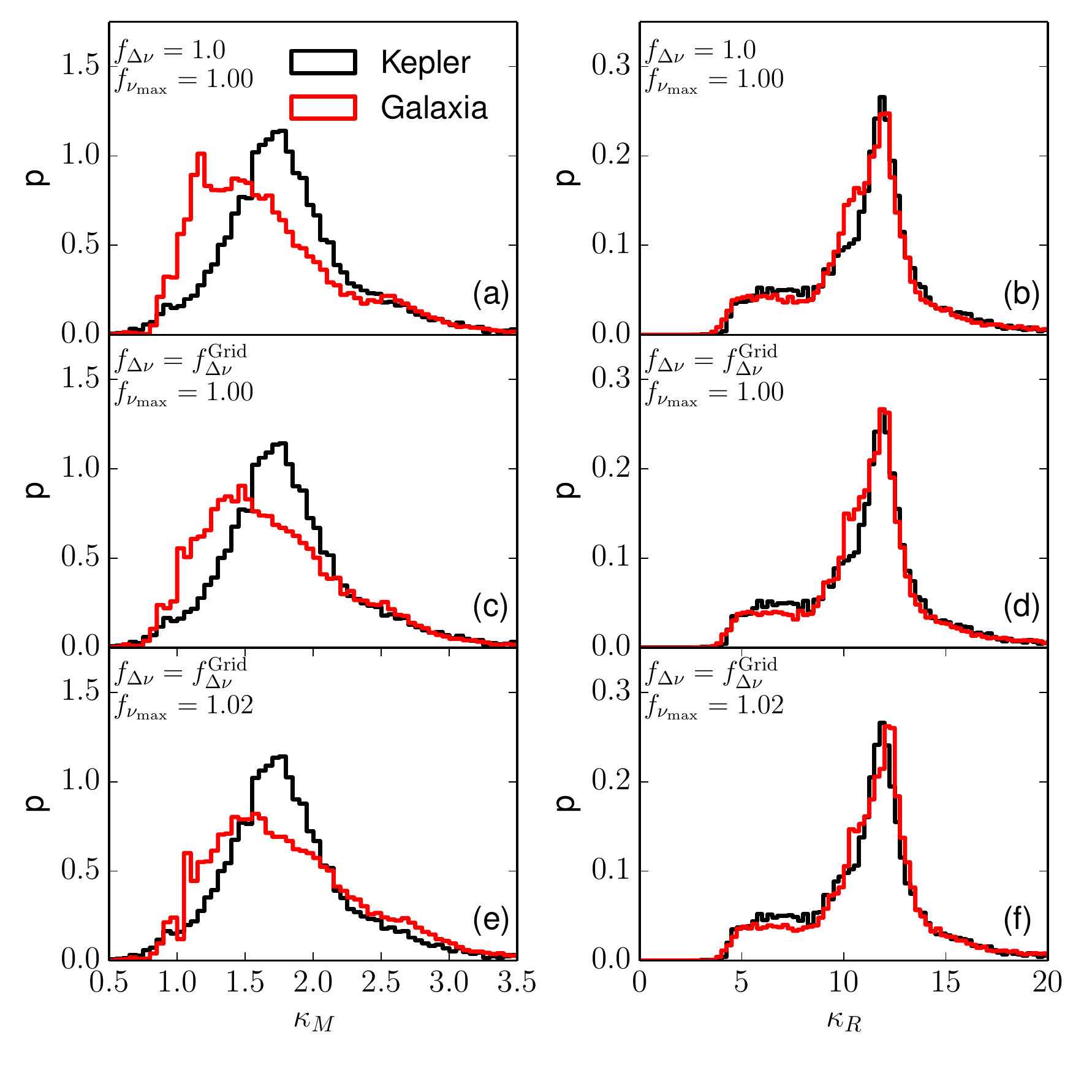}\caption{Comparison of $\kappa_M$ and $\kappa_R$
(Equations \ref{equ:scaling_m_proxy} and
\ref{equ:scaling_r_proxy}) of the {\it
Kepler} red giant sample with the predictions of {\sl
Galaxia}. Each row uses different correction factors to
the scaling relations to
compute $\kappa_M$ and $\kappa_R$ for the synthetic {\sl
Galaxia} sample. When
no corrections are applied
the observed and predicted $\kappa_M$ distributions differ (panel
(a)).
The use  of correction factor $f_{\Delta \nu}^{\rm
Grid}$ (panel (c))  helps to reduce the discrepancy. Setting
$f_{\nu_{\rm max}}=1.02$ further improves the agreement
(panel (e)).
\label{fig:giants_mass}}
\end{figure}

\subsection{Comparison of mass and radius distributions}
The distributions of $\kappa_M$ and $\kappa_R$
(Equations \ref{equ:scaling_m_proxy} and
\ref{equ:scaling_r_proxy}) are shown in \fig{giants_mass}.
In \fig{giants_mass}a,b we do not apply a correction to
either of the scaling relations
($f_{\Delta \nu}=f_{\nu_{\rm    max}}=1$),
while in \fig{giants_mass}c,d we set
$f_{\Delta \nu} = f_{\Delta \nu}^{\rm Grid}$
(obtained from our grid; \sec{massrad}).
In both
cases the $\kappa_{M}$ distributions do not match.
However, the use of $f_{\Delta \nu}^{\rm Grid}$
clearly reduces the
discrepancy.
In \fig{giants_mass}e,f, we apply a correction to
the $\nu_{\rm max}$  scaling relation by setting
$f_{\nu_{\rm max}}=1.02$ to shift
$\kappa_{M}$ to higher values improving the
agreement with observations.
However, the predicted distribution
is still too broad (see Section \ref{sec:saga} and
\ref{sec:revisit} for further discussion on these issues).
We also investigated if choosing higher
photometric uncertainty while generating the SCP
parameters had any effect on our findings (se \sec{scp}).
We found that choosing a higher
photometric uncertainty had no significant effect
on the mass and the radius distribution
of the synthetic stars.

The isochrones in the Galactic model use $Z_{\odot}=0.019$.
However, recent studies
suggest a lower value of $Z_{\odot}\approx 0.014$
\citep{2009ARA&A..47..481A,2011SoPh..268..255C}.
To be consistent with the isochrones in the
Galactic model  we adopted
$Z_{\odot}=0.019$ when computing $f_{\Delta \nu}^{\rm Grid}$
from our grid.
We note that, decreasing $Z_{\odot}$ from 0.019 to 0.012
when computing $f_{\Delta \nu}^{\rm Grid}$ increases the
predicted values of $\kappa_{M}$ by about 6\%. This increase
in $\kappa_{M}$ is similar to the increase in
$\kappa_{M}$ that occurs when $f_{\nu_{\rm max}}$
is increased from 1.0 to 1.02.

\begin{figure}
\centering \includegraphics[width=0.5\textwidth]{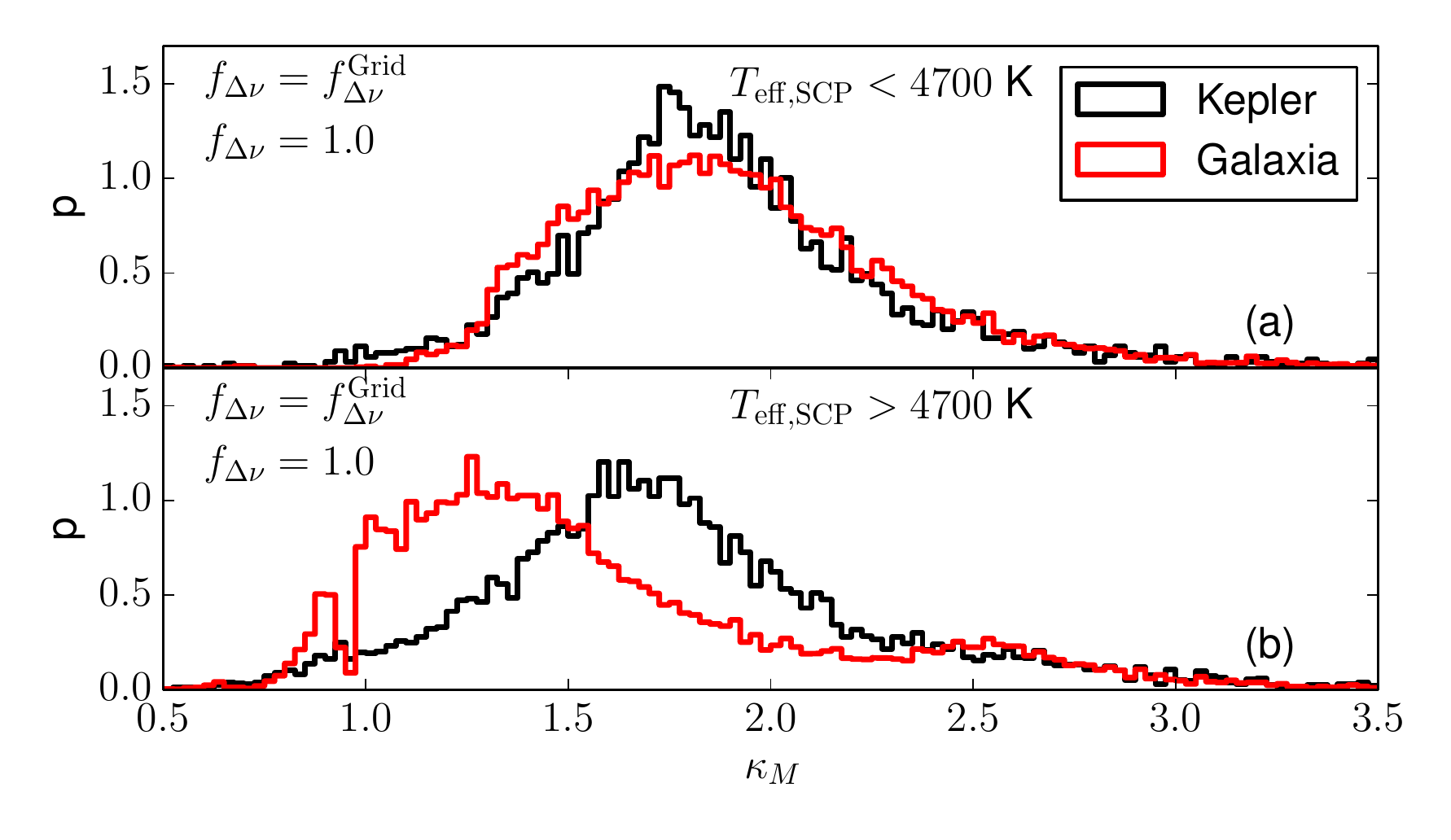}\caption{Comparison of $\kappa_M$  of the {\it
Kepler} red giant sample with the predictions of {\sl
Galaxia} for different temperature limits.
\label{fig:kappam_teff}}
\end{figure}

We next study if the mismatch between
observed and predicted $\kappa_{M}$ distributions has any
dependence on temperature.
We found that the mismatch is
minimal if we restrict the sample to
$T_{\rm eff,SCP}< 4700$ but becomes prominent
with increasing temperature. This is shown in \fig{kappam_teff}
where we plot the $\kappa_{M}$ distributions
for $T_{\rm eff,SCP}< 4700$ and $T_{\rm eff,SCP}> 4700 K$.
$\kappa_{M}$ was computed with $f_{\Delta \nu} =
f_{\Delta \nu}^{\rm Grid}$ and $f_{\nu_{\rm max}}=1.0$.
Also, we use SCP-based
temperatures, because for observed stars we only
have SCP-based temperatures.
The cause behind such a temperature dependent trend is not
yet clear. Firstly, these are SCP-based temperatures, which
may be slightly different from the actual temperatures.
The age, metallicity and temperature are correlated with
each other. So, age and metallicity-based trends
could also manifest themselves as temperature trends.

\begin{figure}
\centering \includegraphics[width=0.45\textwidth]{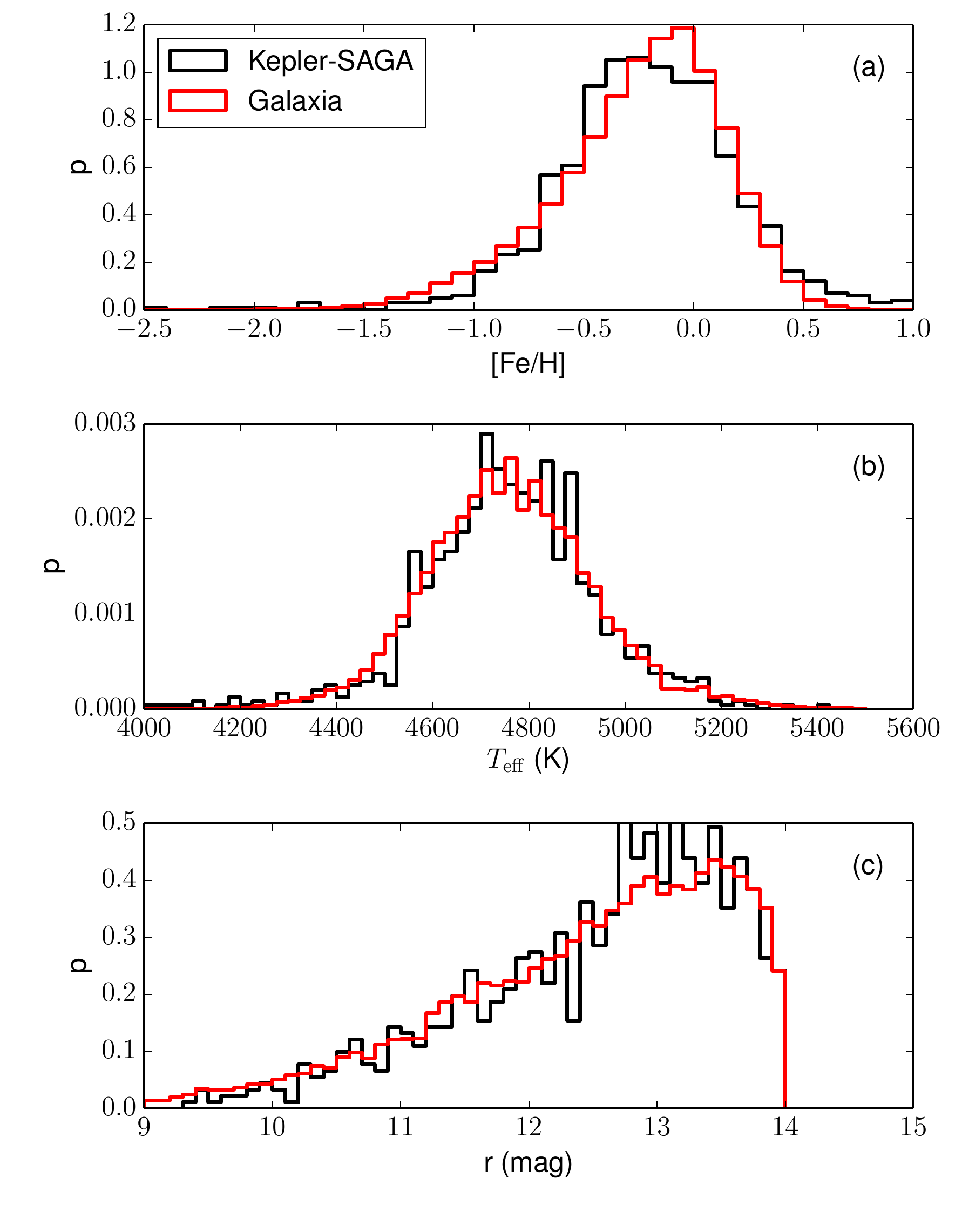}\caption{Comparison of stellar properties in the SAGA survey sample with
the predictions from {\sl Galaxia}. The selection function of
the synthetic stars is the same as used for the {\it Kepler} red giant
sample. An uncertainty of 0.18 dex in ${\rm [Fe/H]}$ was added to
the synthetic stars.
The good match between the distributions
suggest that the SAGA sample is an
unbiased subset of the {\it Kepler} red giant sample.
\label{fig:saga_comp}}
\end{figure}

\subsection{{\it Kepler} giants in the SAGA survey} \label{sec:saga}
In the previous section, we showed that the $\kappa_M$ distribution
of the synthetic stars generated by {\sl Galaxia} does not match
the observed stars. We now investigate if the
mismatch is for all stars in our sample, or whether it depends on
metallicity.
To do this we make use of the SAGA survey, which
provides metallicities for an unbiased subset of red giants
observed by {\it Kepler}.

The SAGA survey \citep{2014ApJ...787..110C} provided
Str{\"o}mgren photometry for giants that have seismic
parameters from {\it Kepler} observations. The current
release comprises, 989 {\it Kepler} red giants based on observations
covering a stripe centered at
galactic longitude $l=74\degree$ and galactic longitude
ranging $7.6<b<19.9$. In SAGA,
Str{\"o}mgren photometry was combined
with broadband photometry, and the infrared
flux method was used to obtain $T_{\rm eff}$ with a precision
of 85 K and ${\rm [Fe/H]}$ with a precision 0.18 dex.
We expect the SAGA sample to be an unbiased subset of the
full {\it Kepler} red giant sample because SAGA observed almost
all stars down to $V=15$ mag, while
the seismic giants are all brighter than 14th mag in $r$
band (roughly corresponding to $V=14.35$).
\citet{2014ApJ...787..110C} reported the SAGA giant
sample to be 95\% complete with respect to the seismic targets.
To check for any potential bias, we
compared the distribution of metallicity, temperature and $r$
band magnitude of the SAGA giants with those of
the synthetic stars satisfying the {\it Kepler} red giant
selection function described in \sec{selfunc} (\fig{saga_comp}).
An uncertainty of 0.18 dex (similar to that of SAGA sample)
was added to the metallicity of the synthetic stars.
The good agreement between the synthetic and the observed
distributions confirms that the SAGA sample is an unbiased
sample of our {\it Kepler} red giant sample.
No other
large unbiased set of metallicities are currently available
for the Kepler red giants.

\begin{figure}
\centering \includegraphics[width=0.45\textwidth]{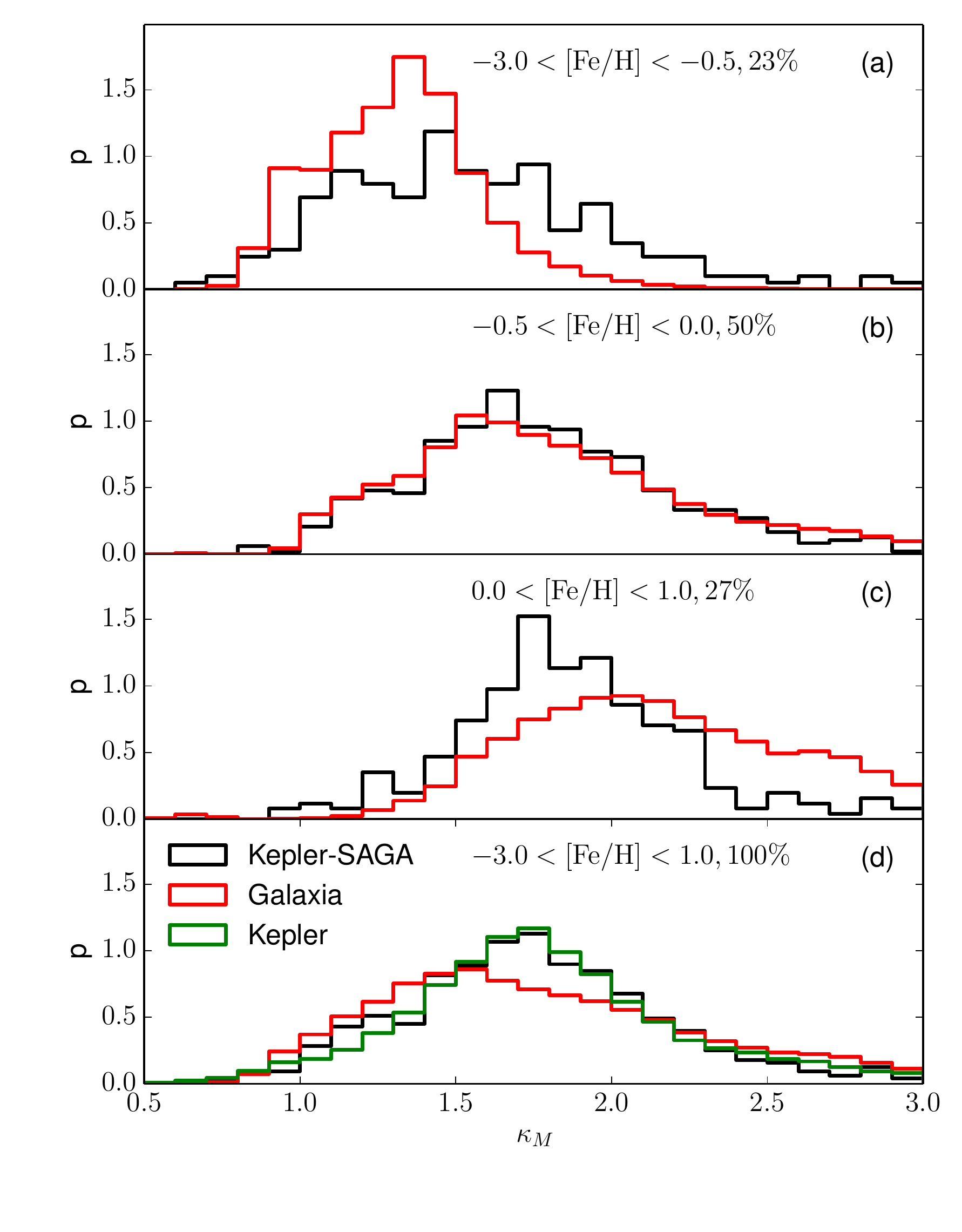}\caption{Comparison of stellar $\kappa_M$ (Equation
\ref{equ:scaling_m_proxy}) from the SAGA
survey sample (black) with
predictions of {\sl Galaxia} (red), using   $f_{\nu_{\rm max}}=1.02$
and $f_{\Delta \nu}=f_{\Delta \nu}^{\rm Grid}$.
The distributions for
different metallicity ranges are split into the different panels. The
sample percentage relative to the full sample
is labelled.
The bottom panel shows the distribution for
the full SAGA sample alongside the distribution
of the full {\it Kepler} red giant sample (green).
\label{fig:saga_mass_fcorr}}
\end{figure}

We now compare the $\kappa_{M}$ distribution  for the SAGA sample
with that from the Galactic model. The
$\kappa_{M}$ values were computed using the correction factor
$f_{\Delta \nu}$ from the grid of stellar models
in \sec{fdnu_corr} and assuming $f_{\nu_{\rm
max}}=1.02$.
\fig{saga_mass_fcorr}a,b,c show that the mean $\kappa_{M}$ increases with
increasing metallicity. However,
the change in $\kappa_{M}$ is stronger in the Galactic model than in
the data. The difference between the model and the data is
largest
for the low and the high metallicity bins, which account for 50\% of the
sample. For $-0.5<\left[{\rm Fe/H}\right]<0.0$, the predictions of the
Galactic model agree with observations.

\fig{saga_mass_fcorr}d shows the distribution of the SAGA
sample for the full ${\rm [Fe/H]}$ range (black) alongside
the full {\it Kepler} red giant sample (green). They
agree with each other, which illustrates once
again that the SAGA sample is an unbiased subset of
the {\it Kepler} red giant sample.

\begin{table*}
\caption{Seismic-inferred stellar properties for benchmark cases where
non-seismic estimates are available}
\begin{tabular}{l l l l l l l l }
\hline
& &  & 1 & 2 & 3 & 4 \\ \hline
benchmark case  & property & non-seismic & seismic &  seismic
& seismic & seismic \\
& & & $f_{\Delta \nu}=1$ & $f_{\Delta \nu}=1$ & $f_{\Delta \nu}=f^{\rm Grid}_{\Delta
\nu}$ & $f_{\Delta \nu}=f^{\rm Grid}_{\Delta  \nu}$ \\
& &  & $f_{\nu_{\rm max}}=1$ & $ f_{\nu_{\rm max}}=1.02$  &
$ f_{\nu_{\rm max}}=1.0$ & $ f_{\nu_{\rm max}}=1.02$  \\
\hline \\
NGC6791, eclipsing binary- & $\langle M_{\rm RGB}\rangle/{\rm M}_{\odot}$ &
$1.15\pm0.02$ & $1.30\pm0.02$ & $1.23\pm0.02$ &
$1.17\pm0.02$ & $1.10\pm0.02$ & \\
extrapolated mass. & $\langle M_{\rm
RC}\rangle/{\rm M}_{\odot}$ & & $1.06\pm0.03$
& $1.00\pm0.03$ & $1.05\pm0.03$ &
$0.99\pm0.03$ &  \\
Brogaard et al. (2012) & & & & & & \\ \\
APOKASC metal poor giants. & $\langle\Delta M \rangle/{\rm M}_{\odot}$ &
$0.0$ &  $0.25\pm0.05$ (OCT) & $0.19\pm0.05$ (OCT)& $0.07\pm0.05$ (OCT)&
$0.017\pm0.05$ (OCT)\\
Epstein et al. (2014) & &
$0.0$ &$0.22\pm0.05$ (SYD)& $0.15\pm0.05$ (SYD)& $0.04\pm0.05$ (SYD)&
$-0.01\pm0.05$(SYD) \\ \\
KIC 8410637, a red giant & $M/{\rm M}_{\odot}$ &
$1.56\pm0.03$  & $1.83\pm0.14$ & $1.73\pm0.14$ & $1.74[1.87]$ \footnote{The
predicted values are assuming the star to be a red giant
branch stars, the values in square brackets are assuming the star
to be a red clump star}$\pm0.14$ & $1.63[1.76]\pm0.14$ \\
in an  eclipsing binary system.& $R/{\rm R}_{\odot}$ &
$10.74\pm0.11$ & $11.58\pm0.3$&
$11.36\pm0.3$ & $11.26[11.71]\pm0.3$ & $11.04[11.48]\pm0.3$ \\
\citet{2013ApJ...767...82G} & $(g/g_{\odot})\times 10^3$ &
$13.52\pm0.4$ & $13.63\pm 0.3$ & $13.36\pm 0.3$ & $13.63\pm 0.3$
& $13.36\pm 0.3$ \\
& $(\rho/\rho_{\odot})\times 10^{3}$ & $1.259\pm0.05$ &
$1.177\pm0.005$& $1.177\pm0.005$ & $1.210[1.164]\pm0.005$ &
$1.210[1.164]\pm0.005$ \\ \\
KIC 9246715, two red giants& $M/{\rm M}_{\odot}$ &
$2.149\pm0.007$ &
$2.30\pm0.08$ & $2.16\pm0.08$ & $2.26[2.37]\pm0.08$ &
$2.12[2.23]\pm0.08$ \\
in an eclipsing binary system. & $R/{\rm R}_{\odot}$ &
$8.30\pm0.05$ & $8.47\pm0.1$&
$8.30\pm0.1$ & $8.40[8.60]\pm0.1$ & $8.23[8.43]\pm0.1$ \\
\citet{2016arXiv160100038R} & $(g/g_{\odot})\times 10^3$ &
$31.19\pm0.4$ & $32.03\pm0.4$ & $31.4\pm0.4$ & $32.03\pm0.4$ & $31.4\pm0.4$  \\
& $(\rho/\rho_{\odot})\times 10^{3}$ & $3.76\pm0.07$ &
$3.78\pm0.02$& $3.78\pm0.02$ & $3.82[3.72]\pm0.02$ &
$3.82[3.72]\pm0.02$ \\ \\
HD185351, a bright giant with &  $R/{\rm R}_{\odot}$ &
$4.97\pm0.07$ & $5.35\pm0.2$ &
$5.24\pm0.2$ & $5.34\pm0.2$   & $5.23\pm0.2$ \\
interferometry and parallax. & & & & & & \\
\citet{2014ApJ...794...15J} & & & & & & \\ \\
\hline
\end{tabular}
\label{tab:mass_conv}
\end{table*}

We note that the analysis of the SAGA sample has its limitations.
The SAGA sample has a metallicity uncertainty of 0.18 dex, which
is relatively large.
So, the suggested range of validity of the Galactic model,
$-0.5<\left[{\rm Fe/H}\right]<0.0$, is subject to this
uncertainty in metallicity.
The good match of $\kappa_M$
in $-0.5<\left[{\rm Fe/H}\right]<0.0$
occurs only
after an ad-hoc correction of 2\% to $\nu_{\rm max}$.
This correction is not a
prescription for a global change to the scaling relation,
but  is a guide to quantify the differences between model
predictions and observations.
We note, it is possible that different groups of
stars (e.g., red clump and red giant branch),  have different
correction factors that negate each other. This would
make such global correction factor unphysical.

\section{Revisiting the asteroseismic scaling relations} \label{sec:revisit}
Our results in the previous sections show that corrections
to the scaling relations can improve the agreement between
observed seismic masses and those predicted by {\sl Galaxia}.
The main correction was to use $f_{\Delta \nu}$
from the grid of stellar models. Additionally, we
investigated a correction of the form $f_{\nu_{\rm
max}}=1.02$, specifically for the red giants.
We now compare the stellar properties like mass and
radius, inferred using seismology with and without
corrections to the scaling relations, against
five benchmark cases where measurements  are
available independent of seismology. The results are
summarized in \tab{mass_conv} and we discuss them in detail below.

\tab{mass_conv} shows that for all five benchmark cases, the
standard uncorrected scaling relations, with $f_{\Delta \nu}=1$ and
$f_{\nu_{\rm max}}=1$ (option 1), overestimate the
mass as compared to non-seismic estimates. We investigate
different combinations of the following
corrections, $f_{\Delta \nu}=f_{\Delta \nu}^{\rm
Grid}$ and   $f_{\nu_{\rm max}}=1.02$ (options 2-4).
We see that  when the $f_{\Delta \nu}^{\rm Grid}$ correction is
used, either with  $f_{\nu_{\rm max}}=1$ or 1.02 (options
3-4), the seismic estimates match the non-seismic estimates
to within $2\sigma$.
Also, when using $f_{\Delta \nu}^{\rm Grid}$, the cluster
NGC 6791 is the only benchmark case that favors
$f_{\nu_{\rm max}}=1$ (option 3) instead of $f_{\nu_{\rm
max}}=1.02$ (option 4).
To conclude, the non seismic
estimates that we study here, support the use of
$f_{\Delta \nu}^{\rm Grid}$. Four out of five benchmark cases
studied also support the use of $f_{\nu_{\rm max}}=1.02$.

\begin{figure}
\centering \includegraphics[width=0.5\textwidth]{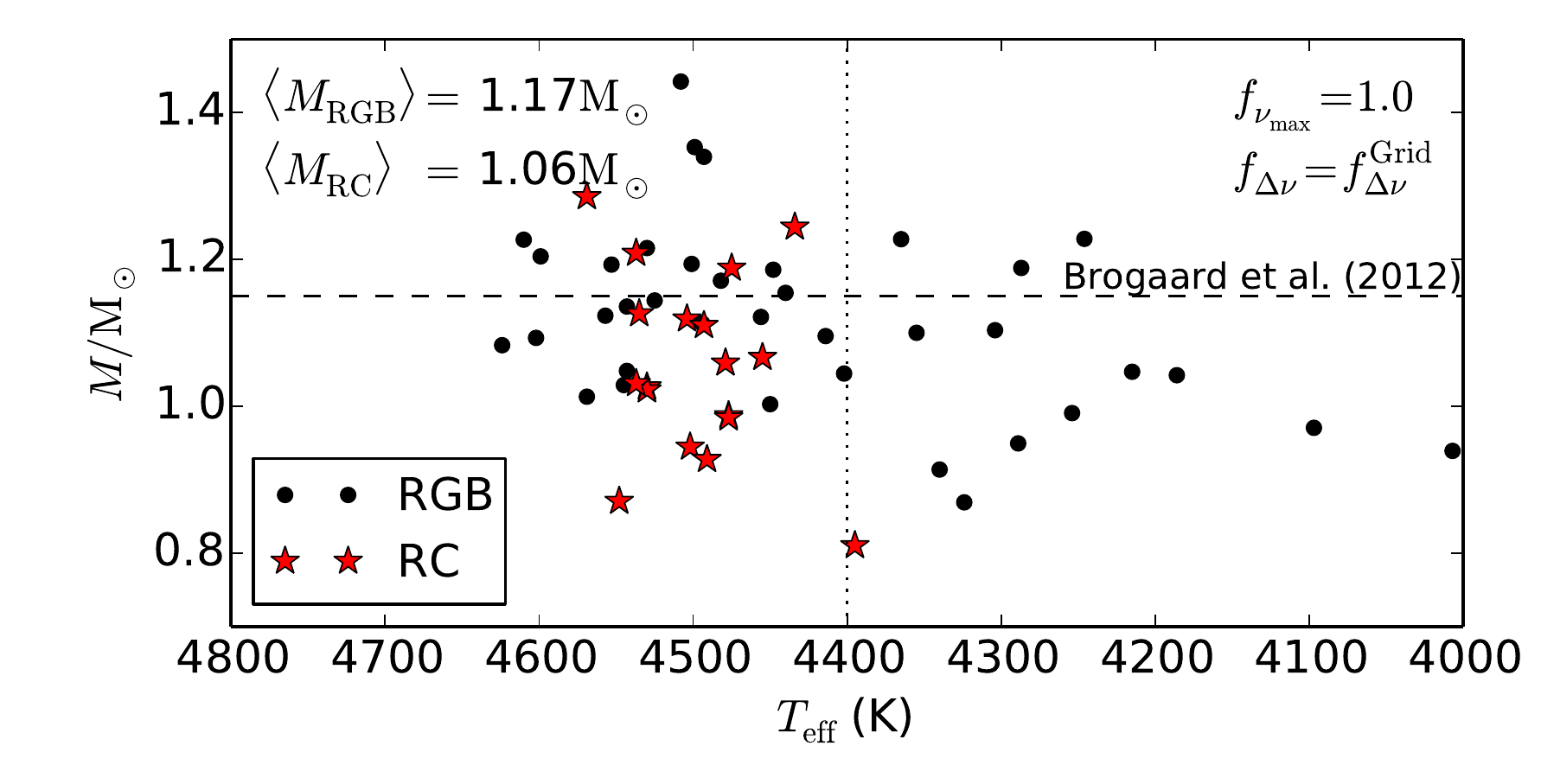}\caption{ Mass of red giant branch and red clump stars in
the cluster NGC 6791 after applying the
correction to the $\Delta \nu$ scaling relation suggested
by stellar models (option 3 from \tab{mass_conv}). The  red clump
stars have a slightly lower
mass on average. The mean mass is computed for stars with $T_{\rm
eff}>4400$ K (left of dotted line) to exclude
asymptotic giant branch and highly evolved red giant
branch stars that might have undergone mass loss.
\label{fig:ngc6791}}
\end{figure}

\begin{table}
\caption{Properties of the open cluster NGC 6791}
\begin{tabular}{l l l}
\hline
& Brogaard 2011,2012 & Miglio 2012 \\
\hline
Metallicity ${\rm [Fe/H]}$  & 0.29  $\pm$0.03$\pm$0.08 &  \\
Distance $(m-M)_V$  & 13.51 $\pm$0.06          &        \\
Helium content $Y$  & 0.3   $\pm$0.01          &        \\
Age                 & 8.3   $\pm$0.4$\pm$0.7   &     \\
$M_{\rm RGB}$         & 1.15  $\pm$0.02          &
1.23$\pm$0.02 \\
$M_{\rm RC}$         &                          & 1.15$\pm$0.03 \\
\hline
\end{tabular}
\label{tab:ngc6791}
\end{table}
\subsection{Masses and radii of stars in the open cluster
NGC 6791}
Clusters provide a unique opportunity to test the scaling
relations because stars in a cluster have a negligible
spread in age, metallicity and distance.
NGC 6791 is one of the oldest and most
metal-rich open clusters known.
\citet{2012A&A...543A.106B} estimated the age, the helium
content, and the mass of stars on the lower red giant branch
using information from two eclipsing binaries
along with color magnitude diagrams of the
cluster \citep[see also][]{2011A&A...525A...2B}. The measured
properties are given in
\tab{ngc6791}.

\citet{2012MNRAS.419.2077M} used
asteroseismology from {\it Kepler}  data to measure the
masses of the red giant branch and red clump stars in NGC6791.
Their estimates of masses for the red giant branch stars
were higher
than found by \citet{2012A&A...543A.106B}. Using photometry,
temperature, and distance modulus, \citet{2012MNRAS.419.2077M}
also measured the radii of the stars independently of asteroseismology.
For the red giant branch stars, they found the radius
estimated from photometry to be
in agreement with those estimated by asteroseismology, but
for the red clump stars they found the seismic estimate
to be smaller by about 5\%.
\citet{2012MNRAS.419.2077M} noted that the
factor $f_{\Delta \nu}$ suggested by their stellar models
was  higher for red clump stars compared
to red giant branch stars.
This prompted
\citet{2012MNRAS.419.2077M} to adopt
$f_{\Delta \nu}=1.0$ for red giant branch stars and $1.027$
for red clump stars, which brought the
seismic and non seismic estimates of radius in agreement
with each other.
When, using $f_{\Delta  \nu}=f_{\Delta  \nu}^{\rm Grid}$ from our stellar models,
we also find $f_{\Delta \nu}$ to be higher for red clump stars
(\fig{teff_dnu_grid}).  However, for NGC6791,
on average the red giant branch stars have
$\langle f_{\Delta \nu} \rangle =0.974$ while the red clump
stars have $\langle f_{\Delta \nu} \rangle =0.999$ (close
to unity).
While the relative difference between the corrections for
the two groups of stars is similar to what
\citet{2012MNRAS.419.2077M} found, our $f_{\Delta \nu}$
values are lower and hence
our seismic estimates of radius, for both red giant branch
and red clump stars,
do not match the non-seismic estimates.
However, a change of
0.15 mag in distance modulus, which has an
uncertainty of 0.06 mag, is enough to bring the
photometric and seismic estimates of radius in agreement.

Our estimates of mass, using $f_{\Delta  \nu}=f_{\Delta
\nu}^{\rm Grid}$,  are shown in \fig{ngc6791}, which
match well with estimates of \citet{2012A&A...543A.106B}
(dashed line).
When estimating the average mass of the red giant branch stars,
we restricted them to be hotter than the coldest red clump star
(hence less luminous than red clump stars) to avoid giants
whose evolutionary state is ambiguous.
\begin{figure}
\centering \includegraphics[width=0.5\textwidth]{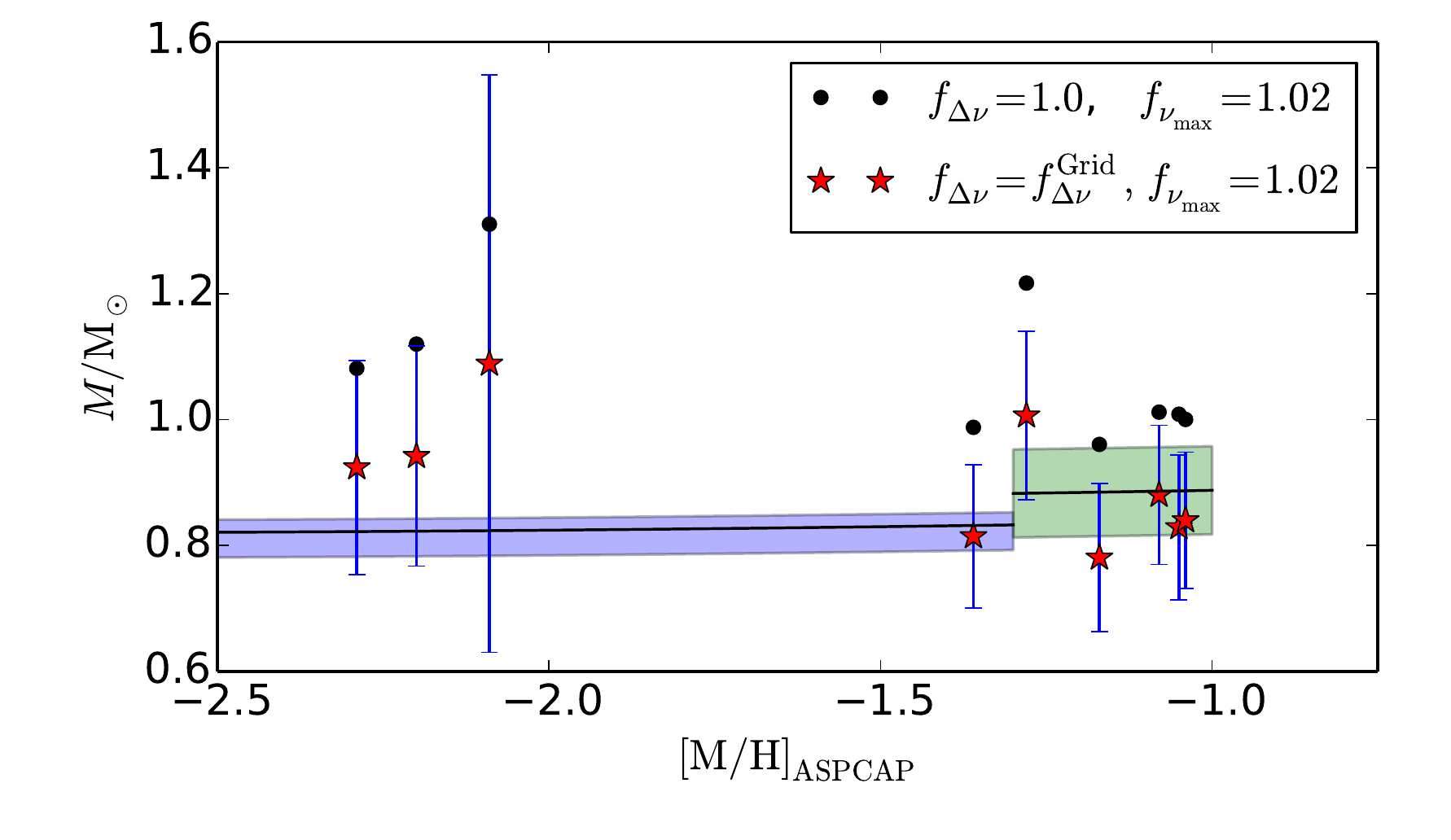}\caption{ Mass of APOGEE metal poor giants selected from
\citet{2014ApJ...785L..28E}. The range of theoretically
expected masses for halo and thick disc stars are shown as
shaded bands and approximately resemble the bands in
\citet{2014ApJ...785L..28E}. The red dots show masses
when correcting scaling relations by $f_{\Delta \nu}=f^{\rm Grid}_{\Delta
\nu}$ and $f_{\nu_{\rm max}}=1.02$. The black dots are masses when no
correction is applied to the $\Delta \nu$ scaling relation and resemble the
estimates of  \citet{2014ApJ...785L..28E} with $\nu_{\rm max,\odot}=3140
\mu$Hz, which is equivalent to $f_{\nu_{\rm max}}=1.016$ in
our terminology.
\label{fig:epstein}}
\end{figure}
\subsection{Metal poor giants in APOKASC}
\citet{2014ApJ...785L..28E} studied asteroseismic masses
of metal-poor stars observed by {\it Kepler} and APOGEE that
were part of the
APOKASC sample \citep{2014ApJS..215...19P}. Nine stars were
found with $[M/H]<-1$. Seven of them were labelled
as halo stars and the other two as thick disc stars, based on their
kinematics. By making some assumptions on $[\alpha/{\rm
Fe}]$ and age of these stars,
\citet{2014ApJ...785L..28E} predicted their masses.
They adopted  a range of
0.2 -- 0.4 for $[\alpha/{\rm    Fe}]$, 10 -- 13.77 Gyr for the age of
the halo and 8 -- 13.77 Gyr for the age of the thick disc.
An approximate reproduction of these predicted masses by us
is shown as the shaded region in \fig{epstein}.
Next, \citet{2014ApJ...785L..28E} measured the mean offset of
seismic masses from these predictions, and found them to
be $\Delta M=0.17\pm0.05 {\rm  M}_{\odot}$.
They used $\nu_{\max,\odot}=3140 \mu{\rm Hz}$, which is
the method-specific value for the OCT method \citep{2010MNRAS.402.2049H}
used by them. This choice is
equivalent to $f_{\nu_{\rm max}}=1.016$ in our terminology
(\equ{scaling_freq2}), because we use $\nu_{\max,\odot}=3090 \mu{\rm Hz}$.
Our results, adopting the OCT values for the stars, with
$f_{\nu_{\rm max}}=1.02$ are shown in  \fig{epstein}
and  \tab{mass_conv}.
The results, both  with
and without $f_{\Delta \nu}^{\rm Grid}$ correction,  are
shown  (options 2 and 4).
We find that adopting the $f_{\Delta \nu}^{\rm Grid}$ correction
lowers the mass on average by 0.15 ${\rm M}_{\odot}$ and brings $\Delta M$
closer to the expected value of zero.
If the OCT method really has method specific systematics
then the above results do not suggest any additional
correction to the $\nu_{\rm max}$ scaling relation besides
that required to acccount for a  higher value of $\nu_{\rm max,\odot}$.
For 8 out of 9 metal poor giants we also have $\Delta \nu$
and $\nu_{\rm max}$ estimates using the SYD pipeline, for
which $\nu_{\max,\odot}=3090 \mu{\rm Hz}$.
Results for these are also shown in \tab{mass_conv}.
Here, option 4 is the best, this supports correction to
both the  $\Delta \nu$ and the $\nu_{\rm max}$ scaling
relations.
A correction to the $\nu_{\rm max}$ scaling relation
was also suggested  by \citet{2014ApJ...785L..28E},
based on two metal poor stars,
$\nu$ Indi \citep{2006ApJ...647..558B}
and KIC 7341231 \citep{2012ApJ...756...19D}.

\subsection{Red giants in  eclipsing binary systems}
Detached eclipsing binaries (dEB) can be used to measure
the mass and radius of each stellar component independently of
asteroseismology, and hence offer the opportunity to
test the asteroseismic scaling relations.
Of the 13 candidate eclipsing
binaries with a pulsating red giant component currently
found in the {\it Kepler} data  \citep{2013ApJ...767...82G},
KIC 8410637 and KIC 9246715 are the only ones that have mass
and radius measured from the analysis of the eclipses in the light curves.
It can be seen from  \tab{mass_conv} that correcting the
scaling relations, with $f_{\Delta \nu}=f^{\rm Grid}_{\Delta
\nu}$ and $f_{\nu_{\rm max}}=1.02$, brings the seismic
masses and radii in better agreement with the dynamical masses
and radii.
The correction factor  $f_{\Delta \nu}$, which depends upon the
evolutionary state of a star, is different for a red
giant branch star and a red clump star.
Currently we do not have any information about the
evolutionary state of these stars.
In \tab{mass_conv}
the values corresponding to a red clump star are shown in
square brackets. It can be seen that
for both the stars the dynamical estimates of mass and radius
favor them to be red giant branch stars
rather than red clump stars.

\subsection{HD185351: radius from interferometry}
Long-baseline interferometry in optical/near-infrared
wavelengths from instruments like CHARA
\citep{2005ApJ...628..453T} allows one to measure angular diameters of stars
with 1-2\% accuracy \citep{2012ApJ...760...32H}. If distance
is known, one can convert angular diameter to radius.
However, accurate reliable distances, e.g.,  from parallax
measurements,  are currently only available for bright nearby stars.
HD185351, with V=5.18, is the third brightest star observed by {\it
Kepler} \citep{2014ApJ...794...15J} and is also nearby
with a parallax-based distance of $40.83\pm0.36$ pc. By combining this distance
with the interferometric
measurement of the angular diameter, \citet{2014ApJ...794...15J}
estimated the radius to be $4.97 \pm 0.07 {\rm R}_{\odot}$.
This estimate is slightly larger than
the asteroseismic estimate of
$5.35 \pm 0.2 {\rm R}_{\odot}$ based on uncorrected scaling
relations. With corrections, $f_{\Delta \nu}=f^{\rm
Grid}_{\Delta  \nu}$ and $f_{\nu_{\rm max}}=1.02$, the
asteroseismic estimate of radius
is $5.23 \pm 0.2 {\rm R}_{\odot}$. which agrees better with
the interferometric and parallax-based measurement.

\begin{figure*}
\centering \includegraphics[width=\textwidth]{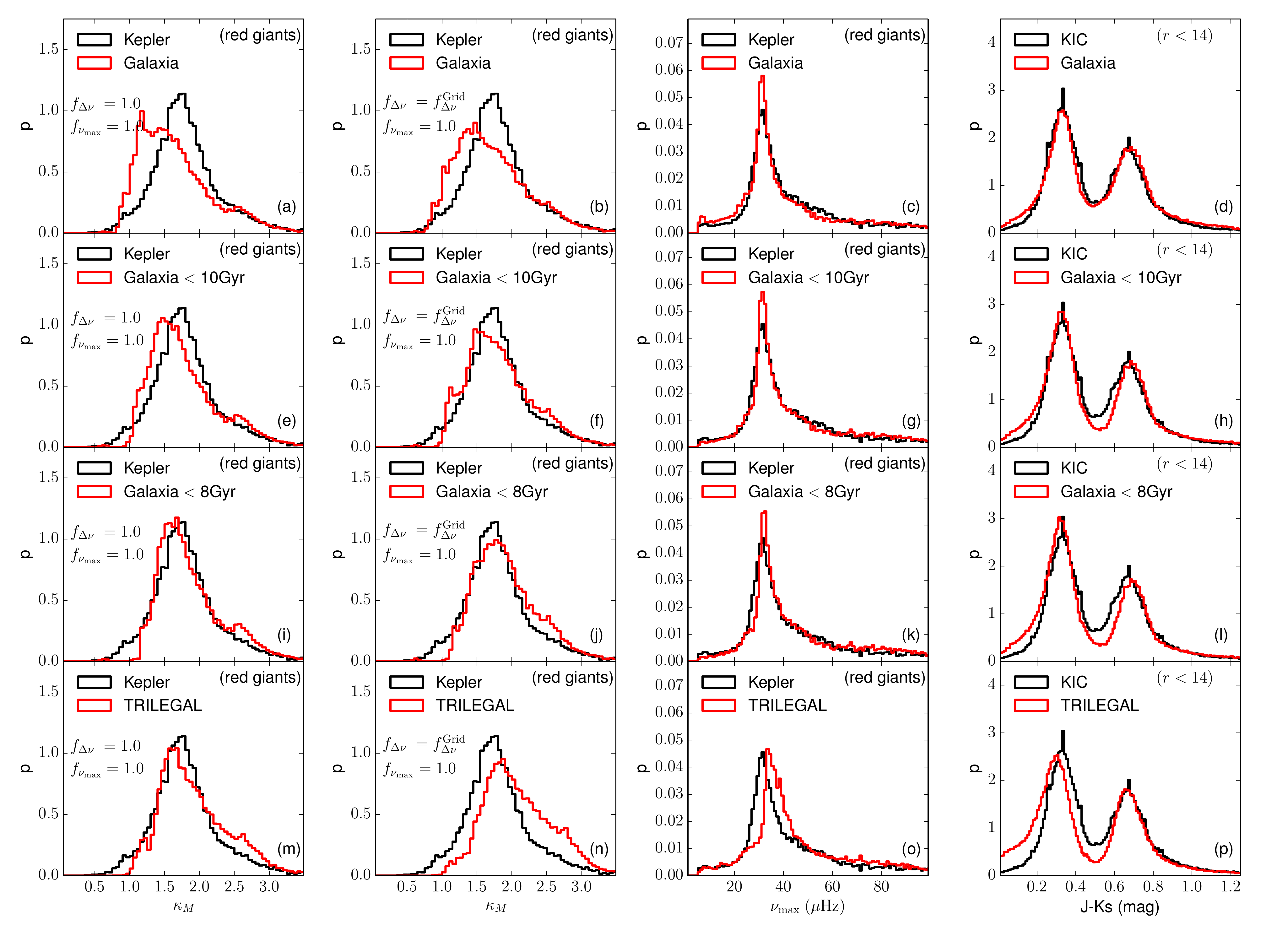}\caption{Comparison between observed and predicted
properties of the {\it Kepler} red giant sample (first
three columns) and a magnitude limited sample from KIC
(last column) for different Galactic
models.  Each row corresponds to a different Galactic model and
from top to bottom these are (a-d)
{\sl Galaxia}, (e-h) {\sl Galaxia} including only stars less
than 10 Gyr old, (i-l) {\sl Galaxia} including only stars less
than 8 Gyr old, and (m-p) TRILEGAL. The $\kappa_M$
distribution in the first
column is computed without any correction to the $\Delta  \nu$ scaling
relation whereas in the second a correction is used.
For the top two rows, a  correction to the $\Delta
\nu$ scaling relation is   required to match the
mass distributions while for the bottom two rows no
correction is required.
The Galactic models that match the observed mass
distribution of the red giant sample do not match the
observed $(J-K_s)$ color
distribution of stars with $r<14$ in KIC.
This is because such models have more younger
stars, which extend the blue wing of the $(J-K_s)$ color
distribution.
\label{fig:galaxia_trilegal_giants}}
\end{figure*}

\section{Changing the Galactic model}
In Section 4, we saw that the predictions
of the {\sl Galaxia} model do not match the  mass distribution
of observed red giant stars from {\it Kepler}. We investigated
corrections to the asteroseismic scaling relations and found that
the corrections improve the agreement with observations but
do not make the disagreements go away completely.
We now investigate the changes that would be required in the Galactic
model to match the asteroseismic information from
{\it  Kepler}. The main mismatch is that the Galactic model
overpredicts the number of low mass stars. The age of a red giant
star is inversely related to its mass, so if we
change the star-formation rate to reduce the relative
fraction of old stars,
the predicted mass distribution should shift towards higher
masses and consequently improve the agreement.

\begin{figure}
\centering \includegraphics[width=0.5\textwidth]{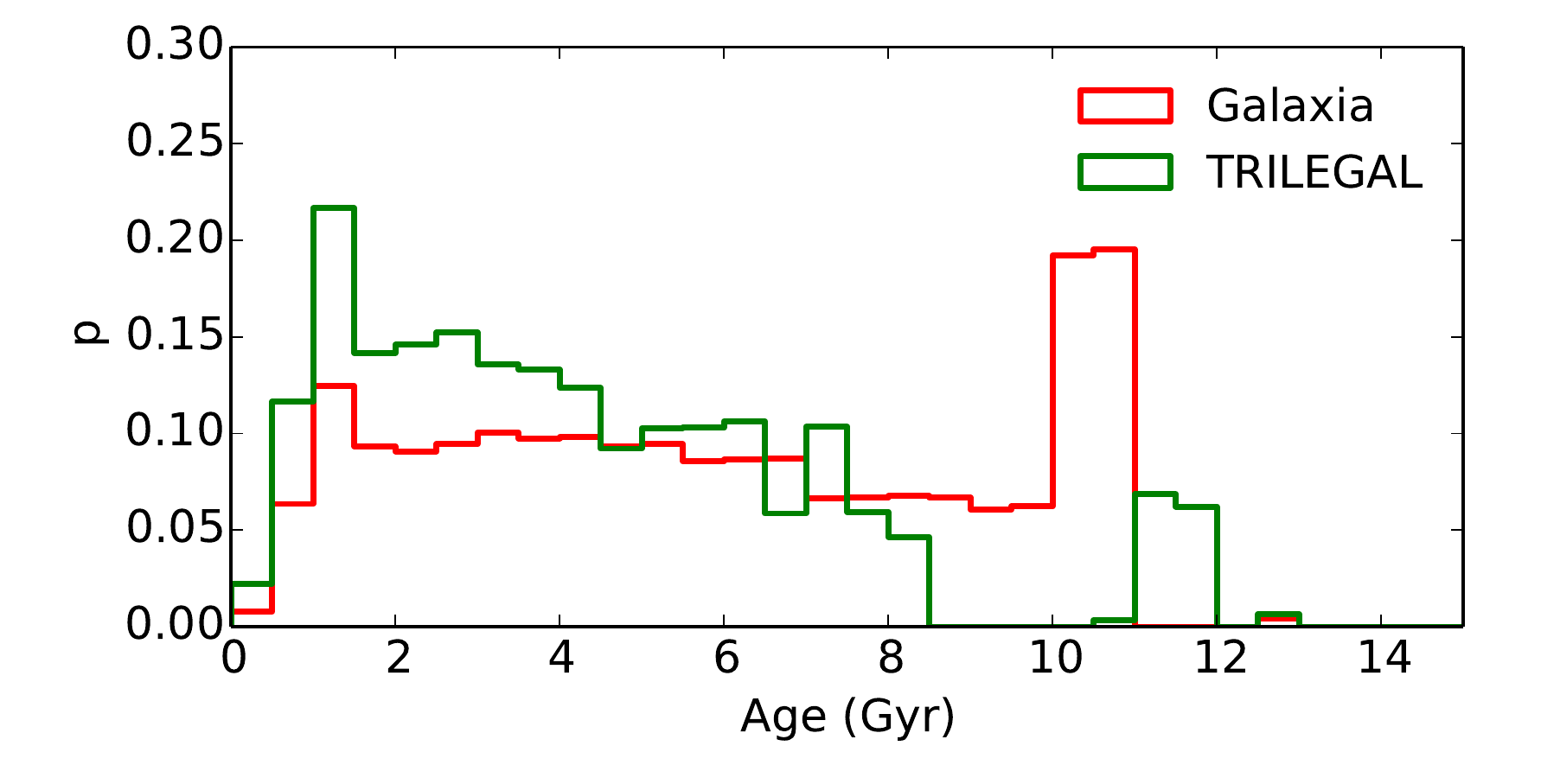}\caption{Age distribution of the {\it Kepler} red giants
as predicted by {\sl Galaxia} and  TRILEGAL. The peak in
the {\sl Galaxia}
distribution between 10-11 Gyr is due to thick disc
stars. In {\sl
Galaxia}, the thin disc has a constant star formation
rate, which
is reflected in the age distribution.
The TRILEGAL model  in comparison has more younger stars and lacks stars older
than 8 Gyr. The thick disc, spanning 11 to 12 Gyr, is also less prominent in
TRILEGAL.
\label{fig:age_dist_galaxia_trilegal_giant}}
\end{figure}

In \fig{galaxia_trilegal_giants}, each row corresponds to
a different Galactic model,
and for each model we plot the
distribution of $\kappa_M$, $\nu_{\rm max}$, and $J-K_S$
color. The distribution of $\kappa_M$, $\nu_{\rm max}$ is
for the red giant sample, while the $J-K_S$ distribution is
for stars in KIC with apparent magnitudes $r<14$.
The $\kappa_M$ distributions are shown both without (first column) and with
correction (second column) to the $\Delta \nu$  scaling
relation.
Panels in the top row (\fig{galaxia_trilegal_giants}a-d)
show the default {\sl Galaxia} model. This has a thin disc
with a constant star
formation rate for age between 0 to 10 Gyr
and a separate thick
disc with age between 10 to 11 Gyr.
In the second row (\fig{galaxia_trilegal_giants}e-h) and the third row
(\fig{galaxia_trilegal_giants}i-l) the star formation rate in {\sl Galaxia} is truncated
such that there are no stars older than 10 and 8 Gyr,
respectively. In the
bottom row (\fig{galaxia_trilegal_giants}m-p) we use the
default version of TRILEGAL, which has
a two step star formation rate. Between 1 to 4 Gyr
the star formation rate is 1.5
times higher than at other times
\citep{2005A&A...436..895G}.
We applied the same selection
procedure on the TRILEGAL stars as we did for {\sl Galaxia}
stars, that is, $g$ band correction, estimation of stellar
parameters from photometry using the SCP code, and then
applying the selection function as discussed in
\sec{selfunc}.
The age distributions of
the red giant sample, as predicted by {\sl Galaxia} and
TRILEGAL, are shown in \fig{age_dist_galaxia_trilegal_giant}. The
TRILEGAL model clearly has more younger stars (in the range 1 --
4 Gyr) than any {\sl Galaxia} based model considered here.
So, going from top row to bottom row,
the percentage of young stars in the model
increases. As expected, the predicted mass distributions,
as shown in the first and second columns, are found
to shift towards higher masses as one
goes from top to bottom along each column.

We now concentrate on the top three rows in
\fig{galaxia_trilegal_giants}, which
are {\sl Galaxia} based models, and
investigate changes relative to the default {\sl
Galaxia} model shown in the first row (\fig{galaxia_trilegal_giants}a-d).
If we remove the very oldest stars (second row/
\fig{galaxia_trilegal_giants}e-h), the
match between the Galactic model and  the
observations is better with the $\Delta \nu$ correction
(panel f) than without (panel e).
Hence, if the correction factor is used,
only stars older than 10 Gyr, which is 19\% of the {\sl
Galaxia} sample,
need to be discarded to bring the models into reasonable
agreement with the data.
If we remove older stars more aggressively (third row/
\fig{galaxia_trilegal_giants}e-h) we see the opposite
effect;  the match is
better when no correction is applied (compare panels i and
j). Hence, if the correction factor is not used,
stars older than 8 Gyr, which is 32\% of the
{\sl Galaxia} sample, need to be discarded to reach a
reasonable match.
These results highlight the degeneracy between
the star formation rate and the $\Delta \nu$ correction factor to the scaling
relations. Since, a correction to the $\nu_{\rm max}$ scaling relation
also alters the mass distribution, the factor $f_{\nu_{\rm
max}}$ will also be degenerate  with the star formation rate.
The bottom row (\fig{galaxia_trilegal_giants}m-p))
shows the TRILEGAL model, whose star formation rate is closer
to the {\sl Galaxia} model in third row
\fig{galaxia_trilegal_giants}i-l). Similar to that
{\sl Galaxia} model, the predicted mass
distribution of TRILEGAL matches with observations,
provided the correction factor
$f_{\Delta \nu}^{\rm Grid}$  is not used.

The models in the bottom three rows of
\fig{galaxia_trilegal_giants} match the observed mass
distribution better than the default {\sl Galaxia} model
shown in the top row (\fig{galaxia_trilegal_giants}a,b).
However, these models
do not match the $(J-Ks)$ color distribution (fourth column),
especially its blue wing where the match becomes progressively
worse
going from top to bottom.
This is because, these models correspond to an increasingly
higher fraction of young stars.
The younger stars are more blue (hotter) giving
rise to an increased number of stars in the blue wing of the
color distribution.
In addition to the mismatch with the observed color
distribution, the TRILEGAL model, also shows a mismatch with the
observed $\nu_{\rm max}$ distribution.
Note that the corrections in \ref{sec:comp_photo} have no effect
on the $J-K_s$ distributions shown here. This is because
any potential inaccuracies in photometry are related to
Kepler $u,g,r,i,$ and $z$ bands and not the 2MASS $J,H,$ and
$K_s$ bands. Also, issues related to the red giant selection
function,  as discussed in \ref{sec:scp}  and
\ref{sec:selfunc}, also have no
effect on the $J-K_s$ distributions shown here, because the
color distribution is shown for a simple
magnitude limited sample satisfying $r<14$.

The star formation rate is not the only part of the model
that can alter the mass distribution.
Changing the age scale height and age scale radius
relation in the Galactic model will alter the age
distribution and thereby the mass distribution
of stars selected with the {\it Kepler} selection function.
Changing the initial
mass function of stars in the Galactic model  can also alter
the mass distribution. Clearly, in future, when attempting to fit a
Galactic model to observed data,
one should take all of the above mentioned factors into account.

\section{Summary and Conclusions}
In this paper, we explored the prospect
of using asteroseismic information of red giants
to constrain stellar population synthesis-based
models of the Milky Way.
We used a sample of about 13,000 red giants from the {\it
Kepler} mission. The large sample size allowed us to test the
Galactic models with asteroseismic data at a level
that was not possible before.
For the first time,
within the framework of stellar
population synthesis-based modelling,
we also investigated changes to the Galactic model and the
asteroseismic relations.

The asteroseismic scaling relations provide a practical way
to determine mass and radius from average asteroseismic
parameters like $\Delta \nu$ and $\nu_{\rm max}$ for a
large number of stars. However, the range of validity
of these relations is not yet clear, specially for red
giants
whose structure is not homologous to that of the Sun.
In previous studies it has been shown that
$\Delta \nu$ estimated from theoretically calculated
oscillation frequencies
differs from the simple $\Delta \nu \propto \rho^{1/2}$
scaling with density. To investigate this, we estimated
corrections to the $\Delta \nu$ scaling relation as a
function of mass, metallicity,
temperature and evolutionary state, using a grid of stellar
models.
We found that the Galactic model
that best matched the photometry
fitted the asteroseismic
data better if the $\Delta \nu$ corrections were
implemented.
We also studied five benchmark cases, where non-seismic estimates
of mass are available, and for all of them the  use of
the $\Delta \nu$ correction brought seismic estimates closer
to non-seismic estimates.
This provides additional support for
the $\Delta \nu$ correction.

Analysis of the SAGA sample, which provides metallicity for
an unbiased subsample of {\it Kepler} giants, allowed us to explore
how the agreement between the Galactic predictions and
observed seismic information depends on metallicity.
We find that the best match of
the {\sl Galaxia} model with observations is
in the range $-0.5<[{\rm Fe/H}]<0.0$.
However, in addition to
correction to the $\Delta \nu$ scaling relation,
a 2\% correction to the $\nu_{\rm max}$ scaling relation is also
required. The correction factor
serves to quantify the difference between theory
and observations and should not be used as a
prescription for changing the seismic scaling relation.
The correction is based purely on comparing the distribution
of masses, and it is possible that red giant branch and red
clump stars have opposite corrections that cancel
each other.

Even after using the correction to the $\Delta \nu$
scaling relation, the Galactic model of {\sl Galaxia}
that best matches the photometric information fails to
match the seismic information for giants.
The mismatch is minimal for $T_{\rm eff}<4700$ but
is significant for $T_{\rm eff}>4700$.
In general, {\sl Galaxia} tends to overestimate the number of low
mass stars, which implies that
{\sl Galaxia} overpredicts the number of old stars.
Further corrections to asteroseismic scaling relations is certainly
a possibility to resolve the discrepancy, as  is a
modification to the Galactic model.
If the scaling relations along with corrections suggested by
stellar models are correct, then at least 20\% of the oldest
stars in {\sl Galaxia} have to be discarded to
bring the predicted mass distributions in
agreement with observations.
However, altering
the star formation rate as suggested above makes the
color distributions disagree with observations.
There are other model parameters that we have
not yet explored, which
can also alter the mass distribution of stars,
such as,  the age-scale height relation, age scale radius relation
and the initial mass function. It is possible that
some combination of these can explain both the photometric
information of stars as well as the asteroseismic
information. This should be explored in future.
Also,
to resolve degeneracies between parameters
governing the model, one would need to explore
observational data from multiple sources, like,
asteroseismic, photometric, spectroscopic and astrometric
missions.

Other than {\sl Galaxia}, we also compared the
predictions of the TRILEGAL Galactic model with
the {\it Kepler} data. While {\sl Galaxia} matches
the photometry of stars in the KIC,
TRILEGAL could not; it overestimated the number of blue stars.
The peak in the $\nu_{\rm max}$ distribution for TRILEGAL
was also shifted compared to the {\it Kepler} observations.
Suppressing older stars in {\sl Galaxia}, also
leads to overestimation of blue stars as seen in TRILEGAL.
This suggests that TRILEGAL overestimates
the occurrence of younger stars. However, TRILEGAL
matches the mass distribution of {\it Kepler} red giants,
provided no corrections  are applied to  $\Delta \nu$.

Although we have been able to derive a selection function
for the Kepler red giant sample that seems to be
representative of the population, we caution
that it might have some inaccuracies
given the complex selection procedure involved in the actual
selection of targets.
Hence, it is important to verify the
current findings using an independent sample of stars
obeying an unambiguous selection function, such as,
the red giants observed by K2 for galactic archaeology
purposes, where stars have been targeted with
a well defined color magnitude limits
\citep{2015ApJ...809L...3S}.

To conclude, our results clearly demonstrate that the
size and quality of asteroseismic data provided by
{\it Kepler} can, in principle, provide good constraints
on Galactic model parameters. However, further work is required
to validate the accuracy of asteroseismic scaling
relation-based
mass estimates for giants and to explore a wider variety of
changes to the Galactic
model input. Missions
like K2, TESS, and PLATO are going to increase the seismic sample
size by two orders of magnitude. There are also plans to get high
resolution spectroscopic data for many of these stars, which will
help to explore the connection between age and
chemistry. Such data,
when combined with distances from Gaia, will be invaluable
for modelling the Milky Way and trying to
understand its formation.

\section*{Acknowledgments}
We acknowledge the support of {\it Galactic
Archaeology and Precision Stellar Astrophysics} program
organized by  Kavli Institute for Theoretical Physics
(National Science Foundation Grant No. NSF PHY11-25915)
for facilitating helpful discussions of results in this
paper; especially Mark Pinsonneault, Andrea Miglio,
Victor Silva Aguirre and Aldo Serenelli. SS is funded 
through Australian
Research Council (ARC) DP grant 120104562 (PI
Bland-Hawthorn) which supports the HERMES project.  
JBH is funded through Laureate Fellowship from the ARC.  
D.H. acknowledges support by the ARC's DP grant DE140101364 
and support by the NASA under Grant
NNX14AB92G issued through the Kepler Participating Scientist
Program.

\bibliographystyle{apj}

\end{document}